\documentclass{aa}
\usepackage{amsmath}
\usepackage{graphicx}
\usepackage{breqn} 
\usepackage[mathscr]{euscript} 
\usepackage{comment}
\usepackage[utf8]{inputenc}

\usepackage{booktabs}
\usepackage{adjustbox}
\usepackage{tablefootnote}

\usepackage{caption}
\usepackage{subcaption}
\usepackage{ulem}

\usepackage{multicol}

\usepackage{multirow}

\usepackage{natbib} 
\bibpunct{(}{)}{;}{a}{}{,} 

\usepackage{txfonts}
\graphicspath{{Graphics/}}

\makeatletter
\renewcommand*\aa@pageof{, page \thepage{} of \pageref*{LastPage}}
\makeatother

\usepackage[]{hyperref}
\hypersetup{unicode=true, colorlinks=true, linkcolor=[rgb]{0.53, 0.15, 0.34}, citecolor=[rgb]{0.0, 0.27, 0.42}, filecolor=[rgb]{1.0, 0.13, 0.32}, urlcolor=[rgb]{0.53, 0.15, 0.34}}

\begin{document}

   \title{The remarkable microquasar S26: a super-Eddington PeVatron?}
   \author{Leandro Abaroa\inst{1,2}\thanks{leandroabaroa@gmail.com}, 
          Gustavo E. Romero\inst{1,2}, Giulio C. Mancuso\inst{1,2}
           \and Florencia N. Rizzo\inst{2}
           }

   \offprints{Leandro Abaroa}
  \institute{Instituto Argentino de Radioastronom\'ia (CCT La Plata, CONICET; CICPBA; UNLP), C.C.5, (1894) Villa Elisa, Argentina \and Facultad de Ciencias Astron\'omicas y Geof\'{\i}sicas, Universidad Nacional de La Plata, B1900FWA La Plata, Argentina}

   \date{Received / Accepted}


\abstract
{S26 is an extragalactic microquasar with the most powerful jets ever discovered. They have a kinetic luminosity of $L_{\rm j}\sim5\times 10^{40}\,{\rm erg\,s^{-1}}$. This implies that the accretion power to the black hole should be super-Eddington, of the order of $L_{\rm acc}\sim L_{\rm j}$. However, the observed X-ray flux of this system indicates an apparent very sub-Eddington accretion luminosity of $L_{\rm X}\approx 10^{37}\,{\rm erg\,s^{-1}}$.}
{We aim to characterize the nature of S26, explain the system emission, and study the feasibility of super-Eddington microquasars as potential PeVatron sources.}
{We first analyze multi-epoch X-ray observations of S26 obtained with \textit{XMM-Newton} and model the super-Eddington disk and its wind. We then develop a jet model and study the particle acceleration and radiative processes that occur in shocks generated near the base of the jet and in its terminal region.} 
{We find that the discrepancy between the jet and the apparent disk luminosities in S26 is caused by the complete absorption of the disk radiation by the wind ejected from the super-Eddington disk. The non-thermal X-rays are produced near the base of the jet, and the thermal X-rays are emitted in the terminal regions. The radio emission observed with the Australia Telescope Compact Array can be explained as synchrotron radiation produced at the reverse shock in the lobes. We also find that S26 can accelerate protons to PeV energies in both the inner jet and the lobes. The protons accelerated in the lobes of S26 are injected into the interstellar medium with a total power of $\sim 10^{36}\,{\rm erg\,s^{-1}}$.}
{We conclude that S26 is a super-Eddington microquasar with a dense disk-driven wind that obscures the X-ray emission from the inner disk, and that the supercritical nature of the system allows the acceleration of cosmic rays to PeV energies.}

\keywords{cosmic rays --  relativistic processes -- X-ray: binaries -- gamma-rays: general -- radiation mechanism: nonthermal}
\authorrunning{L. Abaroa et al.}
\titlerunning{The microquasar S26: a super-Eddington PeVatron?}

\maketitle
\section{Introduction} \label{sec: intro}



Microquasars (MQs) are binary systems in which a compact object (such as a black hole or neutron star) draws matter from its companion star, leading to the formation of an accretion disk. Bipolar and highly collimated outflows emerge from the central region of these sources. Since the observation of the first MQs three decades ago \citep{1994Natur.371...46M}, many of these systems have been discovered and studied, exhibiting a wide variety of characteristics \citep[e.g.][]{1999ARA&A..37..409M,2005A&A...429..267B,2008A&A...485..623R,2010LNP...794.....B,2015ApJ...807L...8B,2017MNRAS.472.4220B,2017SSRv..207....5R,2017MNRAS.471.3657Z,2018A&A...612A..27K,2021MNRAS.506.1045M,2024MNRAS.528L.157D}.

S26 is a unique microquasar located in the Sculptor Galaxy NGC 7793 at a distance of 3.9 Mpc \citep{2003A&A...404...93K}, whose nebula has a size of $\sim 350 \times 185\,{\rm pc}$ \citep{2002ApJ...565..966P,dopita2012}. According to observations in the optical/X-ray band, and following the relation of \cite{1977ApJ...218..377W} between the wind-driven stellar bubbles and the jet-driven radio lobes, \cite{2010Natur.466..209P} showed that this MQ has the most powerful jets in accreting binaries, with a mechanical luminosity of $L_{\rm j}\sim 5\times 10^{40}\,{\rm erg\,s^{-1}}$. These authors also estimated its age $(t\sim 2\times10^5\,{\rm yr})$, 
proposed that the binary system at the core of S26 is composed of a black hole and an early Wolf-Rayet (WNE) star, and identified nonthermal X-rays produced near the black hole. \cite{soria2010} used the Australia Telescope Compact Array (ATCA) to resolve the radio lobe structure of S26, and suggested that the radio emission from the terminal region of the jets is consistent with synchrotron mechanisms. They also analyzed X-ray observations from the \textit{Chandra} space telescope, identifying X-ray hot spots $\approx20\,{\rm pc}$ farther out than the peak of radio intensity in the lobes, and argued that this emission is most likely of optically thin thermal origin. Other authors have constrained the properties of the expanding nebula of S26 by spectroscopic analysis \citep{dopita2012} and studied its high-energy and neutrino emission \citep{2017APh....90...14I}. 

A key parameter for characterizing MQs is the accretion rate of matter onto the compact object. Accretion proceeds in three basic regimes, depending on the ratio of the actual accretion rate to the Eddington rate. At very sub-Eddington rates, the inner disk is optically thin and geometrically thick, while at moderate rates the regime approaches the standard Shakura-Sunyaev disk, which is optically thick and geometrically thin. In super-accreting black holes \citep{1980ApJ...242..772A,2004PASJ...56..569F}, the radiation pressure overcomes gravity in the innermost part of the disk, and its surface layers are ejected as a consequence of the self-regulation of accretion at the Eddington rate \citep{2009PASJ...61.1305F}. In these systems, the accretion disk launches powerful winds with mass loss rates similar to the accretion rate \citep{Abaroa_etal_2023}. An example of a super-Eddington MQ in our Galaxy is SS433, whose jets have a kinetic power of $\approx 10^{39}\,{\rm erg\, s^{-1}}$ \citep[see][for a detailed review]{2004ASPRv..12....1F} and presents opaque equatorial outflows \citep[e.g.,][]{Picchi_etal2020}. Following the jet-disk symbiosis model \citep{1995A&A...293..665F}, the kinetic luminosity of the jet of S26 estimated by \cite{2010Natur.466..209P} implies that the accretion luminosity to the stellar-mass black hole should be super-Eddington, on the order of $L_{\rm acc}\sim L_{\rm j}$. However, the observed X-ray flux from the core of the system, measured by the \textit{Chandra} and \textit{XMM-Newton} telescopes, indicates an apparent very sub-Eddington luminosity of $L_{\rm X}\sim 10^{37}\,{\rm erg\,s^{-1}}$, orders of magnitude smaller than the jet power.

In this paper we study the properties of S26 and propose an explanation for the apparent discrepancy between the disk and jet luminosities. We hypothesize that the compact object is a 10 solar mass black hole accreting at super-Eddington rates, and that its actual accretion power is a few times $10^{40}\,{\rm erg\, s^{-1}}$. We suggest that some energy should also be extracted from the ergosphere of the spinning black hole to provide enough power for the jet \citep{1977MNRAS.179..433B}. In such a scenario, the efficiency of converting accretion power into jet power can exceed 100 percent in highly rotating black holes, reaching transient values of 300 percent in the most extreme cases \citep{2010ApJ...711...50T,2011MNRAS.418L..79T,2012MNRAS.423.3083M}. On the other hand, as mentioned above, one of the main features of these supercritical systems is a very dense wind ejected from the accretion disk. 

In our hypothesis, the opaque wind will almost completely obscure the disk, making it impossible for an observer to see the X-rays produced in its innermost region. If this is the case, then the X-ray flux detected from the core would not be coming from the disk, but from some other region of the system. This emission could be produced at the base of the jet by internal shocks, close to the black hole but above the photosphere of the disk-driven wind. The phenomenology of S26 is essentially similar to that of an ultraluminous X-ray source (ULX), but seen from a large inclination angle: if the viewing angle were close to $0^{\circ}$, the observer would probably be able to see the innermost disk and thus the thermal hard X-ray component with high luminosity, enhanced by the geometric beaming \citep{2001ApJ...552L.109K,2009MNRAS.393L..41K,2010MNRAS.402.1516K,Kaaret_2012_ulx,Fabrika_etal_2012_ULX,Abaroa_etal_2023, Combi_etal_2024,Veledina_etal_cygx3_2024NatAs}.

We present here a study of the thermal and nonthermal radiation of S26 and we explore the relationship between the jet and the disk in this source. We analyze multi-epoch X-ray observations of the system obtained with \textit{XMM-Newton} and develop a jet model to explain the emission. We study particle acceleration and radiative processes that occur in adiabatic shocks generated at the base of the jet and in the lobes. We also examine the feasibility of microquasars as potential PeVatron sources, exploring their ability to accelerate protons to energies of about 1 PeV or higher. 

We first describe the observations made by the \textit{XMM-Newton} space observatory and the results of our data analysis. Then, we describe in \hyperref[sect: physical model]{Sect.~\ref{sect: physical model}} our model for the accretion disk, its wind, and the jets. We detail all the thermal and nonthermal radiative processes of the system in \hyperref[sect: radiative processes]{Sect.~\ref{sect: radiative processes}}. We present the results of our calculations in \hyperref[sect: results]{Sect.~\ref{sect: results}} and, after a discussion (\hyperref[sect: discussion]{Sect.~\ref{sect: discussion}}), we conclude with a summary and our conclusions.

\section{X-ray observations}

The \textit{XMM-Newton} Observatory (\citealt{jansen2001}) consists of three X-ray telescopes, each equipped with a European Photon Imaging Camera (EPIC) at its focus. The EPIC instrument includes the PN detector (\citealt{struder2001}) and two MOS detectors (\citealt{turner2001}) covering the 0.15--15 keV energy range with a spectral resolution of $R = E/{\rm d}E \approx 20$--$50$.

S26 was within the field of view of \textit{XMM-Newton} on 22 occasions between May 2012 and December 2023, located approximately 3--4 arcmin from the center point (on axis). Since S26 is an extragalactic source, to maximize the number of photons and thus the signal-to-noise ratio, we specifically selected long-duration observations, excluding those with raw exposure times less than 40~ks from the initial dataset. Therefore, we focused our analysis on 11 observations meeting this criterion.\footnote{There are two additional observations of S26 with raw exposure times greater than 40~ks, but they are not currently public, so they were not included in the analysis presented in this paper.}

We reduced and analyzed the \textit{XMM-Newton} data using the Science Analysis System (\textsc{SAS}) software package version 20.0.0 and the High Energy Astrophysics Software (\textsc{HEASoft}). We calibrated, filtered, and cleaned the event lists by applying the \texttt{epproc} and \texttt{emproc} tasks, for PN and MOS respectively, using the most up-to-date calibration files. 
We then identified and removed periods dominated by high particle background, defined as count rate greater than 0.35 ${\rm cts\,s^{-1}}$ for MOS and count rate greater than 0.4 ${\rm cts\,s^{-1}}$ for PN.

To extract the source spectrum, we selected a circular region of 15 arcsec radius centered at $\alpha$=23$^{\rm h}$58$^{\rm m}$00$^{\rm s}$, $\delta$=--32°33'21'' (J2000; \citealt{dopita2012}). To account for the background emission, photons were selected from a source-free 30 arcsec circular region on the same chip as the source.
We used the \texttt{evselect} task to retain source events for spectra with FLAG == 0 and single and double events with PATTERN <= 4 for the PN camera, and PATTERN <= 12 for the MOS cameras.
We generated the response matrix file (rmf) and ancillary response file (arf) using the  \texttt{rmfgen} and \texttt{arfgen} tasks, respectively. 
%



Finally, we further refined our dataset by excluding from our analysis the observations that were dominated by background above 2~keV, as this provided insufficient data for a robust spectral analysis of the source.
We then analyzed three observations with obsIDs 0748390901 (October 12, 2014), 0804670301 (May 5, 2017), and 0853981001 (November 22, 2019), all performed using the EPIC full frame mode. The average durations for these observations were 46.1, 46.9, and 37.3~ksec, respectively. The list of observations used in this study, along with their corresponding net exposure times after filtering, is presented in \hyperref[tab:Tabla_obsIDs]{Table~\ref{tab:Tabla_obsIDs}}.

\begin{table} 
\centering
\caption{Details of the \textit{XMM-Newton} observations of S26.}
\label{tab:Tabla_obsIDs}
\begin{adjustbox}{max width=\columnwidth}
\begin{tabular}{l c c c c}
\hline
\hline
\rule{0pt}{2.5ex} ObsID & Date & \multicolumn{3}{c}{Filtered Exposure Time (ks)} \\
\rule{0pt}{2.5ex} & mm/dd/yy & PN & MOS1 & MOS2 \\
\hline 
\rule{0pt}{2.5ex} 0748390901 & 12/10/2014 & 42 & 48.2 & 48.2 \\ 
\rule{0pt}{2.5ex} 0804670301 & 05/05/2017 & 31.6 & 54.5 & 54.5 \\
\rule{0pt}{2.5ex} 0853981001 & 11/22/2019 & 27.8 & 42.1 & 41.8 \\
\hline 
\end{tabular}
\end{adjustbox}
\end{table}

\begin{figure}[ht!]
  \centering
  \begin{minipage}{0.4\textwidth}
    \centering
    \includegraphics[width=9cm, height=7cm]{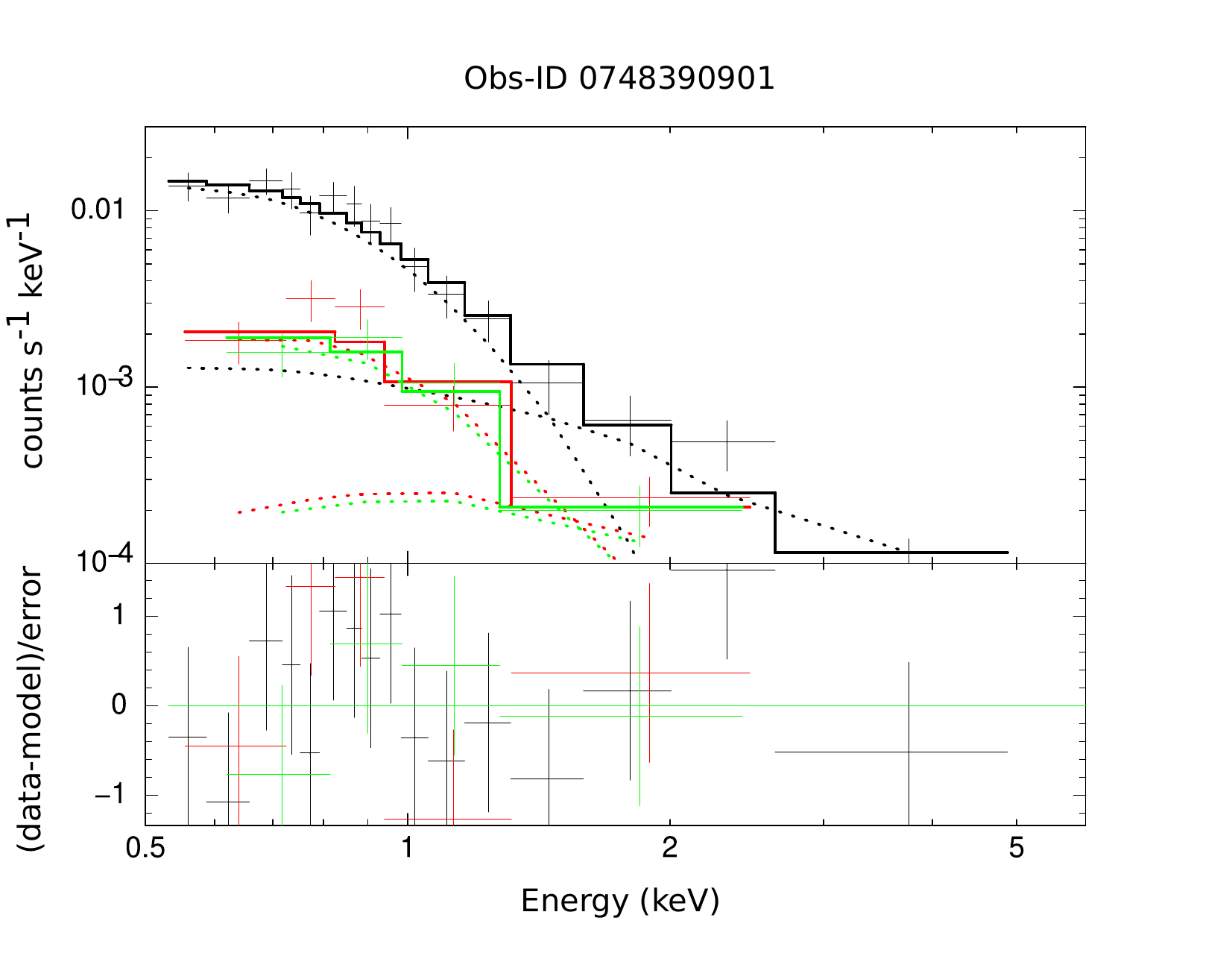}
  \end{minipage}
  \hfill
  \begin{minipage}{0.4\textwidth}
    \centering
    \includegraphics[width=9cm, height=7cm]{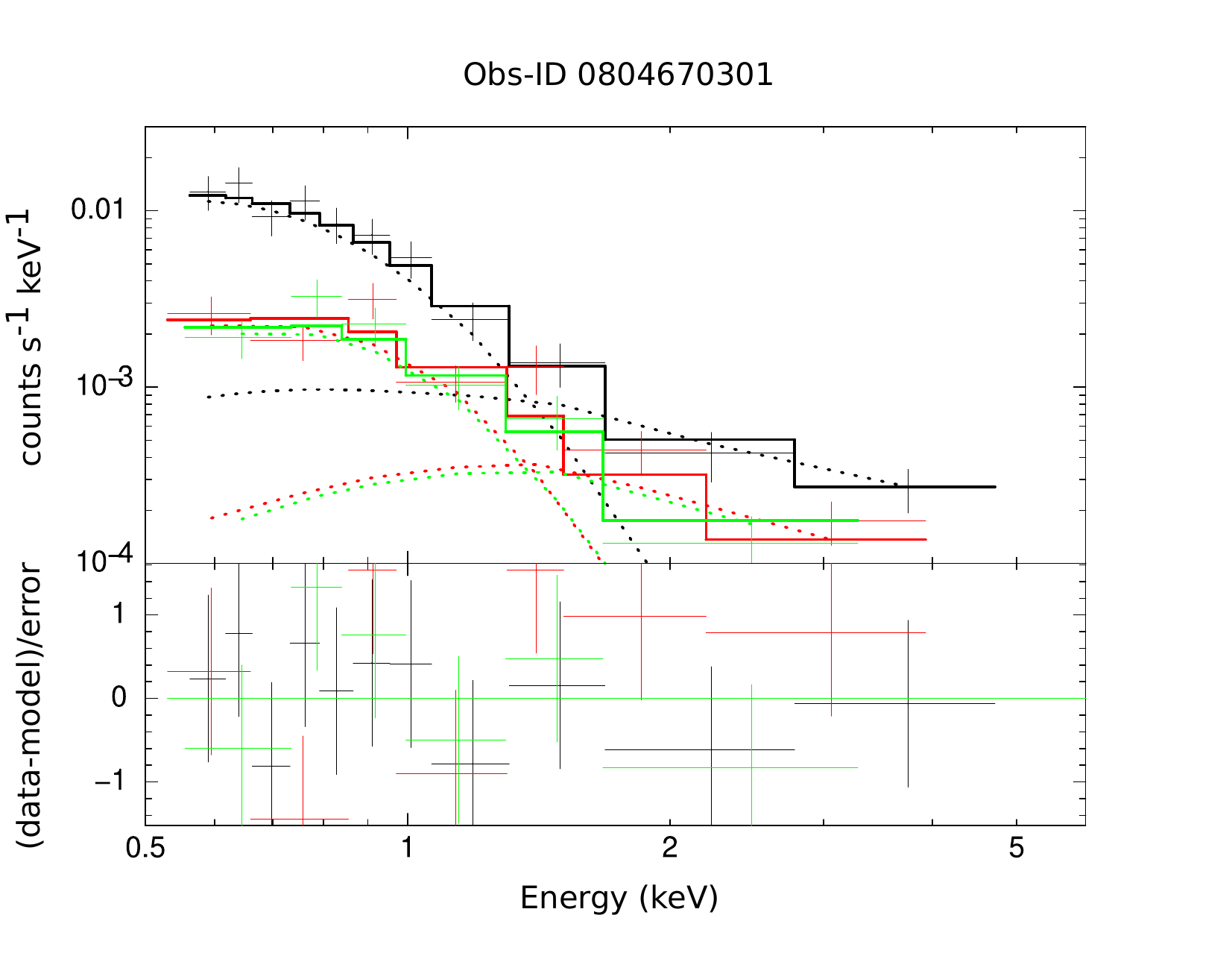}
  \end{minipage}
  \hfill
  \begin{minipage}{0.4\textwidth}
    \centering
    \includegraphics[width=9cm, height=7cm]{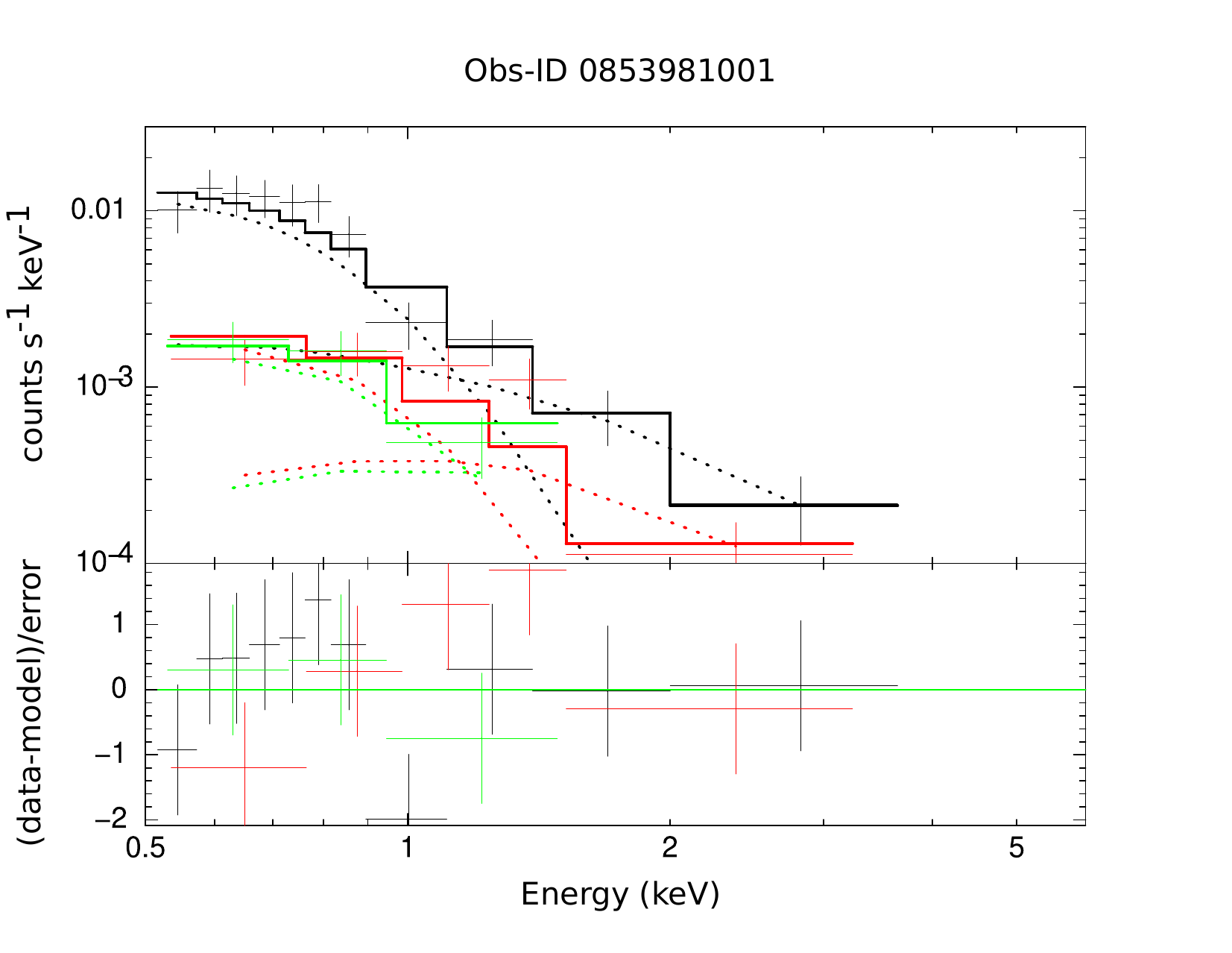}
  \end{minipage}
  \caption{XSPEC \textsc{constant*tbabs*(bbody+powerlaw)} fitting model of the spectrum of S26 using observations 0748390901 (top), 0804670301 (middle) and 0853981001 (bottom) taken by \textit{XMM-Newton}. The black, red, and green colors represent the PN, MOS1, and MOS2 cameras, respectively.}
  \label{fig:fitting_models}
\end{figure}

\subsection{Spectral analysis} 

We performed the X-ray spectral analysis using the \textsc{XSPEC} software v.20.0.0 (\citealt{arnaud1996}) in the 0.5--\textbf{5}.0 keV energy range for obsIDs 0748390901 and 0804670301, and 0.5--4.0 keV for obsID~0853981001.

We attempted different binning for the spectra using the \texttt{specgroup} tool. We applied a minimum of 16 counts per bin for obsIDs 0748390901 and 0853981001, and 20 counts per bin for obsID 0804670301 to allow for Gaussian statistics when fitting, and with an oversampling factor of 3.

We began by fitting the spectra of each observation with two different single-component models. First, we fit the three \textit{XMM-Newton}/EPIC spectra simultaneously with both an absorbed power-law (\texttt{const*tbabs*powerlaw} in XSPEC) and an absorbed blackbody model (\texttt{const*tbabs*bbody} in XSPEC).

Because of the lack of sufficient detected photons, we fixed the value of the hydrogen column density in all cases to $N_{\rm H} = 1.2 \times 10^{20}$ cm$^{-2}$ \citep{kalberla2005}. We adopted the \cite{wilms2000} abundance tables and cross sections from \cite{verner1996}. We fixed the constant factor to unity for the PN camera, but left it free to vary for the two MOS cameras to account for uncertainties in the cross-calibration. We also constrained the model parameters between the three detectors.
Neither model provided a formally acceptable fit; the blackbody model did not adequately account for the observed emission above 2 keV, while the power-law model resulted in a poorly constrained spectral index.

%
%


We then applied a model consisting of a power-law and a blackbody component affected by photoelectric absorption (\texttt{const*tbabs*(powerlaw + bbody}) in XSPEC). We again fixed the model parameters, except for $N_{\rm H}$, which was frozen, and the constant factor, which was fixed at one for PN but allowed to vary freely for the MOS cameras. This model gave a good fit to the observed emission, with $\chi^{2} = 16.44/16.38/16.45$ for 19/18/13 degrees of freedom (dof).

Overall, the parameters obtained with the PN and MOS cameras are consistent with each other within errors. 
The best-fitting parameters are reported in \hyperref[tab:Tabla_model_parameters]{Table~\ref{tab:Tabla_model_parameters}} (the errors indicate 1$\sigma$ confidence levels), while the corresponding spectra and fitted model are shown in \hyperref[fig:fitting_models]{Fig.~\ref{fig:fitting_models}}.

The model fit yielded a power-law photon index $\Gamma$ and a blackbody temperature $kT_{\rm bb}$ which are consistent with the values reported in previous X-ray studies \citep{soria2010}. 
We obtained a 0.5--5.0 keV unabsorbed flux of $1.8^{+0.2}_{-0.3} \times 10^{-14}$ $\rm erg\, cm^{-2}\, s^{-1}$ and $2.8^{+0.3}_{-0.4} \times 10^{-14}$ $\rm erg\, cm^{-2}\, s^{-1}$ for obsIDs 0748390901 and 0804670301, respectively, and for obsID 0853981001, a 0.5--4.0 keV unabsorbed flux of $1.7^{+0.1}_{-0.7} \times 10^{-14}$ $\rm erg\, cm^{-2}\, s^{-1}$. Assuming a distance of $d = 3.9$ Mpc \citep{2003A&A...404...93K} these fluxes correspond to an unabsorbed luminosity of the order of $\sim 10^{37}$ $\rm erg\, s^{-1}$.
The X-ray spectra of S26 obtained with {\it XMM-Newton} cannot be spatially resolved due to the limited resolution of the instrument. Therefore, we used the smallest possible extraction region to obtain the spectrum that includes the entire source. Thus, our model suggests that the emission from S26 could have two possible origins, one thermal and one nonthermal. This is consistent with the results reported by \citet{soria2010}, who showed that the X-ray spectrum of S26 can be modeled by a combination of a thermal component and a nonthermal power law, the former associated with the lobes and the latter associated with the emission from the core. These observations were made with the \textit{Chandra} space observatory, which has a better spatial resolution.


The temperature of the thermal component has negligible errors and remains essentially constant on timescales of years between observations (with a confidence of $\sim 81\%$). Conversely, the photon index of the nonthermal component has such large errors that we cannot rule out variability in the X-ray emission from the core.

\begin{table*}
\begin{center}
\caption{Fitting parameters for the X-ray emission of S26 using the \textsc{constant*tbabs*(bbody+powerlaw)} model.}
\label{tab:Tabla_model_parameters}
\begin{adjustbox}{max width=\textwidth}
\begin{tabular}{l c c c c c c c c c c c}
\hline
\hline
 &  &  &  &  & Model &  &  &  &  &  \\
\hline
\rule{0pt}{2.5ex} ObsID & Constant & Constant & Constant & TBabs & Bbody & Bbody & Powerlaw & Powerlaw & Flux & Chi Squared \\
\rule{0pt}{2.5ex} & PN & MOS1 & MOS2 & $N_{\rm H}$ & $kT_{\rm bb}$ & normalization & $\Gamma$ & normalization & 0.5--E$_{\rm max}$ keV& \\
\hline
\rule{0pt}{2.5ex} 0748390901 & 1 (fixed) & 0.9$^{+0.1}_{-0.1}$ & $0.8^{+0.1}_{-0.1}$ & $1.2 \times 10^{20}$ (fixed) & 0.16$^{+0.01}_{-0.01}$ & $\left(2.7^{+0.3}_{-0.4}\right)\times 10^{-7}$  & 1.4$^{+0.7}_{-0.7}$ & $\left(1.2^{+1.0}_{-0.7}\right)\times 10^{-6}$ & $1.9_{-0.3}^{+0.2} \times 10^{-14}$ & 16.44 (19 dof) \\

\rule{0pt}{2.5ex} 0804670301 & 1 (fixed) & 0.9$^{+0.1}_{-0.1}$ & 0.8$^{+0.1}_{-0.1}$ & $1.2 \times 10^{20}$ (fixed) & 0.16$^{+0.01}_{-0.01}$ & $\left(3.1_{-0.4}^{+0.4}\right)\times 10^{-7}$ & 0.8 $_{-0.5}^{+0.5}$ & $\left(1.6_{-0.8}^{+1.0}\right)\times 10^{-6}$ & 3.1$_{-0.7}^{+0.1} \times 10^{-14}$ & 16.38 (18 dof) \\

\rule{0pt}{2.5ex} 0853981001 & 1 (fixed) & 1.0$_{-0.2}^{+0.2}$ & 0.9$_{-0.2}^{+0.2}$ & $1.2 \times 10^{20}$ (fixed) & 0.13$_{-0.02}^{+0.02}$ & $\left(2.6_{-0.9}^{+0.5}\right)\times 10^{-7}$ & 1.6$_{-1.2}^{+1.0}$ & $\left(1.9_{-1.3}^{+1.8}\right) \times 10^{-6}$ & 1.7$_{-0.3}^{+0.1} \times 10^{-14}$ & 16.45 (13 dof) \\

\hline
\end{tabular}
\end{adjustbox}
\end{center}
\end{table*}


\section{Physical model of S26} \label{sect: physical model}

\begin{figure*}
    \centering
    \includegraphics[width=17cm]{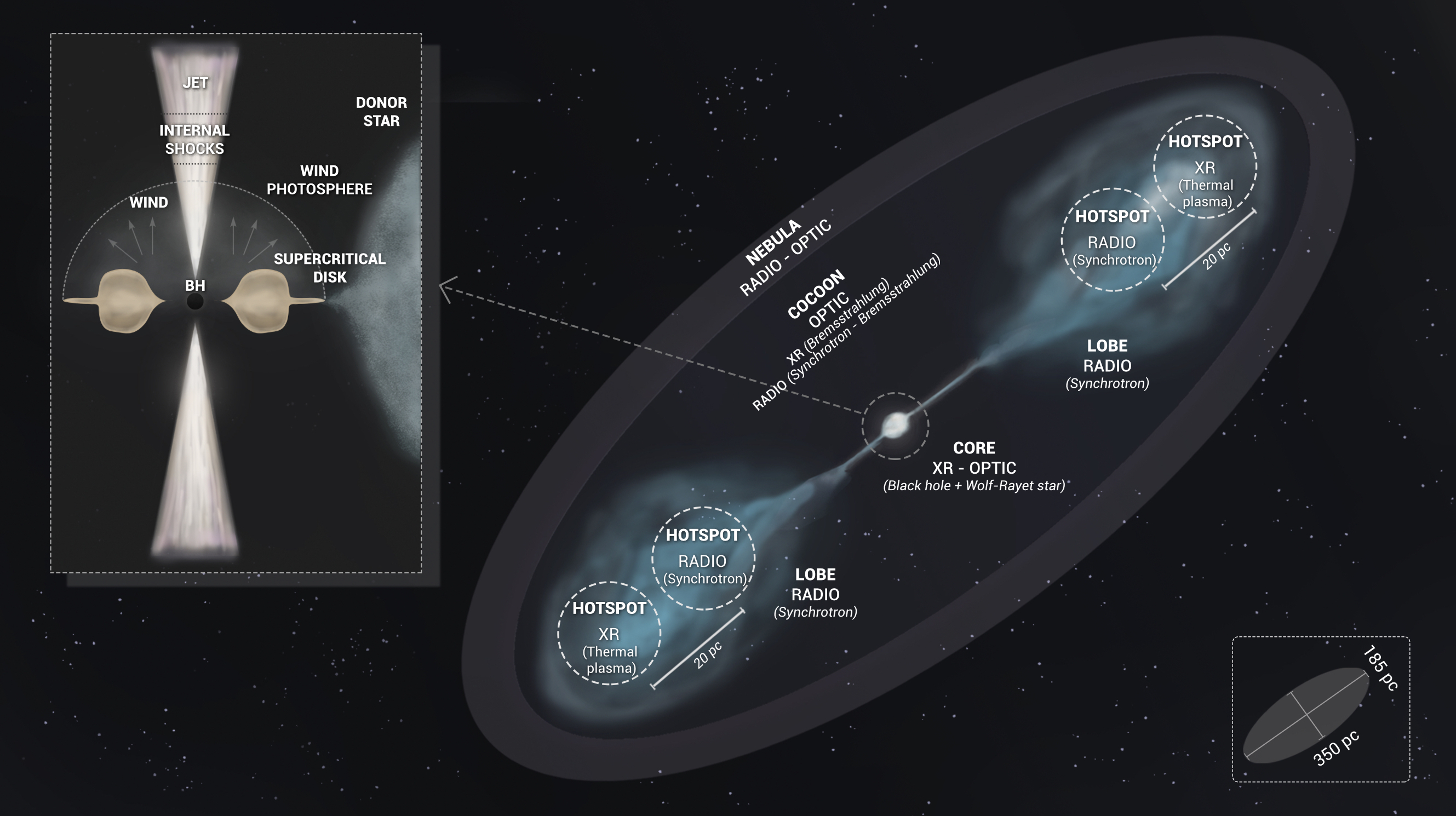}
    \caption{Conceptual scheme of the microquasar S26, not to scale. The stellar-mass black hole and the Wolf-Rayet star are in the core. The extended lobes are at the ends of the two opposite jets, where shocks form at the hotspots. There, the backward shock accelerates particles to relativistic energies, producing synchrotron radio emission, and the forward radiative shock produces optically thin thermal X-ray emission. nonthermal X-ray emission is produced by the electron-synchrotron mechanism at the base of the jet in the core, close to the black hole but above the photosphere of the disk-driven wind.}
    \label{fig: scheme}
\end{figure*}

We assume that the X-ray binary consists of a black hole (BH) with mass $M_{\rm BH}=10\,M_{\odot}$ and an early-type donor star. The star overflows its Roche lobe, transfers mass to the BH through the Lagrange point, and an accretion disk is formed due to the angular momentum of the system. In the following, we describe the semi-analytical models we use to study the accretion disk of the BH, the wind ejected from its surface, and the jet. We show a schematic picture of S26 according to our model in \hyperref[fig: scheme]{Fig.~\ref{fig: scheme}} and list all the parameters in \hyperref[tab: parametros generales]{Table~\ref{tab: parametros generales}}. 

\begin{table} 
\begin{center}
\caption{Parameters of our physical model of S26.}
\label{tab: parametros generales}
\begin{adjustbox}{max width=\columnwidth}
\begin{tabular}{l c c c}
\hline
\hline
\rule{0pt}{2.5ex}Parameter & Symbol & Value & Units  \\
\hline
\rule{0pt}{2.5ex}System \\  
\hline 
\rule{0pt}{2.5ex}Distance to the source$^{(3)}$ & $d$ & 3.9 & ${\rm Mpc}$ \\
Age of the microquasar$^{(4)}$ & $t$ & $2\times 10^5$ & ${\rm yr}$ \\
Total nebula size$^{(5)}$ & & $350\times 185$ & ${\rm pc}$ \\
\hline
\rule{0pt}{2.5ex}Black hole\\  
\hline
\rule{0pt}{2.5ex}Mass$^{(1)}$ & $M_{\rm BH}$    & 10 & $M_{\odot}$ \\
Gravitational radius$^{(2)}$ & $r_{\rm g}$   & $1.50\times 10^6$  & $\rm{cm}$ \\
Spin$^{(2)}$ & $a$   & $>0.80$  & \\
\hline
\rule{0pt}{2.5ex}Accretion disk\\
\hline
\rule{0pt}{2.5ex}Eddington accretion rate$^{(2)}$ & $\dot{M}_{\rm Edd}$ & $2.20\times 10^{-7}$ & $M_{\odot} \ \rm{yr}^{-1}$ \\
Accretion input of matter$^{(1)}$ & $\dot{M}_{\rm input}$ & $2.20\times 10^{-6}$ & $M_{\odot} \ \rm{yr}^{-1}$ \\
Critical radius$^{(2)}$ & $r_{\rm crit}$ & $5.90\times 10^8$ &  ${\rm cm}$ \\
Magnetically-saturated radius$^{(1)}$ & $r_{\rm ms}$ & $1.50\times 10^8$ &  ${\rm cm}$ \\
Accretion power at $r_{\rm ms}^{(2)}$ & $L_{\rm accr}(r_{\rm ms})$  & $3.15\times10^{40}$ & ${\rm cm\,{s}^{-1}}$\\
\hline
\rule{0pt}{2.5ex}Wind \\  
\hline 
\rule{0pt}{2.5ex}Mass loss in winds$^{(2)}$ & $\dot{M}_{\rm w}$ & $2\times10^{-6}$ & $M_{\odot} \ \rm{yr}^{-1}$ \\
Velocity$^{(2)}$ & $v_{\rm w}$ & $1.5\times 10^9$ & ${\rm cm\,s^{-1}}$ 
\\
Temperature of the photosphere$^{(2)}$ & $T_{\rm photo}$ & $3\times 10^5$ & ${\rm K}$ 
\\
Height of the photosphere$^{(2)}$ & $z_{\rm photo}$ & $2.60\times10^{9}$ & ${\rm cm}$ \\
\hline
\rule{0pt}{2.5ex}Jet \\ 
\hline
\rule{0pt}{2.5ex}Mechanical power$^{(4)}$ & $L_{\rm j}$  & $5\times10^{40}$ & ${\rm erg\,{s}^{-1}}$\\
Semi opening angle$^{(1)}$ & $\theta_{\rm j}$  & $5.70$  & degrees \\
Inclination $^{(1)}$ & $i$  & $75$ & degrees \\
Accretion-to-jet efficiency transfer $^{(2)}$ & $q_{\rm j}$  & $1.60$ & \\
Jet bulk velocity $^{(1)}$ & $v_{\rm j}$  & $2.80\times10^{10}$ & ${\rm cm\,{s}^{-1}}$\\
Normalized velocity$^{(1)}$ & $\beta_{\rm j}$  & $0.90$ & \\
Lorentz factor$^{(1)}$ & $\gamma_{\rm j}$  & $3$ & \\
Launching point $^{(1)}$ & $z_{\rm 0}$  & $1.50\times10^{8}$ & ${\rm cm}$\\
Magnetic field near the BH $^{(2)}$ & $B_0$   & $1.60\times10^{9}$ & ${\rm G}$  \\
\hline
\rule{0pt}{2.5ex}Jet base \\ 
\hline
\rule{0pt}{2.5ex}Location of acceleration region$^{(1)}$ & $z_{\rm b}$  & $5\times10^{9}$ & ${\rm cm}$\\
Acceleration region size$^{(1)}$ & $\Delta x_{\rm b}$  & $5\times10^{8}$ & ${\rm cm}$  \\
Acceleration efficiency$^{(2)}$ & $\eta_{\rm acc}$  & $0.10$ & \\
Power to relativistic particles$^{(1)}$ & $q_{\rm rel}$  & $0.01$ & \\
Magnetic field at acceleration point$^{(2)}$ & $B_{\rm b}$   & $2.60\times10^{5}$ & ${\rm G}$  \\
Hadron-to-lepton ratio$^{(1)}$ & $K_{\rm b}$  & $1$ & \\
Spectral index$^{(1)}$ & $p_{\rm b}$  & $2$ & \\
Number density of protons$^{(2)}$ & $n_{\rm b}$  & $5\times10^{9}$  & ${\rm cm^{-3}}$\\
Hillas energy$^{(2)}$ & $E_{\rm Hillas}$  & $2\times 10^{17}$  & ${\rm eV}$\\
\hline
\rule{0pt}{2.5ex}Jet lobe\\ 
\hline
\rule{0pt}{2.5ex}Location of acceleration region$^{(1)}$ & $z_{\rm l}$  & $4\times10^{20}$ & ${\rm cm}$  \\
Acceleration region size$^{(1)}$ & $\Delta x_{\rm l}$  & $4\times10^{19}$ & ${\rm cm}$  \\
Acceleration efficiency$^{(2)}$ & $\eta_{\rm acc}$  & $0.10$ & \\
Power to relativistic particles$^{(1)}$ & $q_{\rm rel}$  & $0.01$ & \\
Magnetic field at acceleration point$^{(5)}$ & $B_{\rm l}$   & $2.50$ & $\mu{\rm G}$  \\
Reverse shock velocity$^{(4)}$ & $v_{\rm rs}$  & $1.5\times10^{10}$ & ${\rm cm\,{s}^{-1}}$\\
Hadron-to-lepton ratio$^{(1)}$ & $K_{\rm l}$  & $100$ & \\
Spectral index$^{(1)}$ & $p_{\rm l}$  & $1.80$ & \\
Number density of protons$^{(6)}$ & $n_{\rm l}$  & 0.10  & ${\rm cm^{-3}}$\\


\hline
\end{tabular}
\end{adjustbox}
\end{center}
\footnotesize{\textbf{Notes.} We indicate the parameters we have assumed  with superscript ${(1)}$ and those we have derived  with ${(2)}$. The other superscripts indicate that the parameters were taken from \cite{2003A&A...404...93K} (3), \cite{2010Natur.466..209P} (4), \cite{dopita2012} (5), and \cite{soria2010} (6). Subscripts 'b' and 'l' refer to base and lobe, respectively.}
\end{table}

\subsection{Accretion disk}
We assume that the BH accretes matter at a super-Eddington rate at the outer part of the disk, $\dot{m}=\dot{M}_{\rm input}/\dot{M}_{\rm Edd} \gg 1$, where $\dot{M}_{\rm input}$ is the input of mass per unit of time. The Eddington rate is given by
\begin{equation} \label{eq: tasa eddington}
    \Dot{M}_{\rm{Edd}}= \frac{L_{\rm{Edd}}}{\eta c^2} \approx 2.2\times 10^{-8} M_{\rm BH} \, {\rm yr^{-1}} = 1.4 \times 10^{18} \frac{M}{M_\odot} \, \rm{g \, s^{-1}},
\end{equation}
with $L_{\rm Edd}$ the Eddington luminosity (defined as the luminosity required to balance the attractive gravitational pull of the BH by radiation pressure). The parameter $\eta \approx 0.1$ is the accretion efficiency \citep{2004PASJ...56..569F}, and $c$ is, as usual, the speed of light. 

The critical radius, given by $r_{\rm crit} \sim  40 \,\dot{m} \,r_{\rm g}$, separates the disk into two regions: a standard outer disk \citep{1973A&A....24..337S} and a radiation-dominated inner disk with advection \citep{2004PASJ...56..569F}, where $r_{\rm g}=GM_{\rm BH}/c^2$ is the gravitational radius of the BH, with $G$ the gravitational constant. 
The disk becomes geometrically thick in the inner region, where the ejection of winds by the radiation force helps to regulate the mass-accretion rate onto the BH ($\dot{M}_{\rm acc}$) at approximately the Eddington rate \citep{2004PASJ...56..569F}:
\begin{equation}\label{intensidad}
    \dot{M}_{\rm acc}(r_{\rm d})  =
    \left\lbrace \begin{array}{l}
   \dot{M}_{\rm input}, \ \ r_{\rm d} > r_{\rm crit}\\ \\
 \dot{M}_{\rm input}\,\dfrac{r_{\rm d}}{r_{\rm crit}}=\dfrac{1}{40}\dfrac{r_{\rm d}}{r_{\rm g}}\dot{M}_{\rm Edd}, \ \ r_{\rm d} \le r_{\rm crit}
    \end{array} 
    \right.
\end{equation}
where $r_{\rm d}$ is the distance to the black hole in the equatorial plane of the disk. Such regulation leads to a total mass loss rate in the winds $\dot{M}_{\rm w}$ that is approximately equal to the accretion input, $\dot{M}_{\rm w}\approx \dot{M}_{\rm input}\gg \dot{M}_{\rm Edd}$. The accretion power at a radius $r_{\rm d}$ is given by $L_{\rm acc}(r_{\rm d})\approx \dot{M}_{\rm acc}(r_{\rm d})\,c^2$.  

In this supercritical scenario, the X-ray emitting region of the disk is completely obscured by the opaque wind described below. The radiation from the disk is therefore absorbed and reprocessed by the wind and shifted to lower energies.

\subsection{Absorption by the disk-driven wind} \label{sec: wind photosphere}

\begin{figure}[ht]
    \centering    \includegraphics[width=10cm]{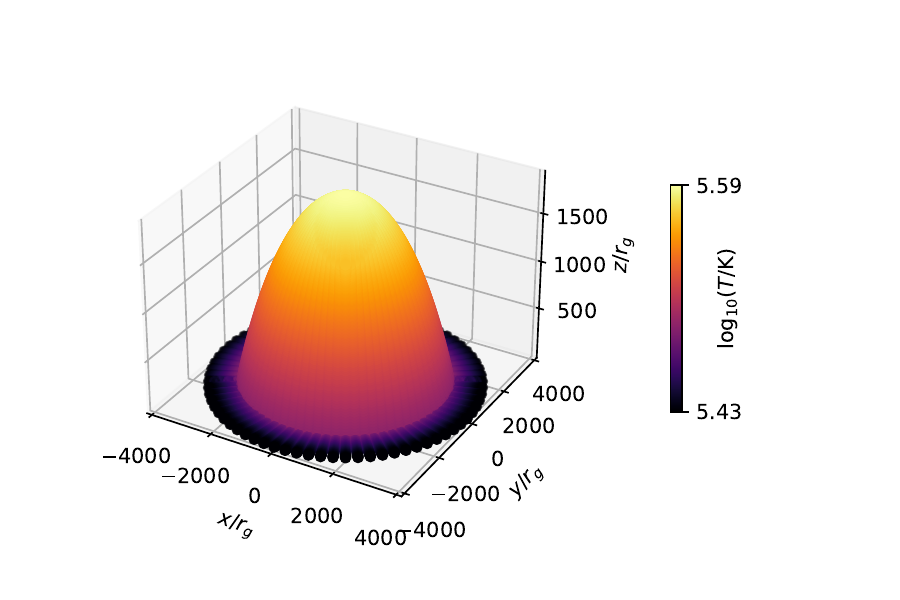}
    \caption{Photosphere of the disk-driven wind for an accretion rate of $10\dot{M}_{\rm Edd}$. The axes are in Cartesian coordinates and units of the gravitational radius. The height $z_{\rm photo}$ is $1777\,r_{\rm g}$ $(\approx 2.6\times10^{9}\,{\rm cm})$ above the BH and its top temperature is $\sim 3.9\times 10^5\,$K.}
    \label{fig: photosphere}
\end{figure}

We assume that the wind ejected from the BH disk is symmetric, smooth, and expands at a constant velocity, estimated to be $\beta_{\rm w}=v_{\rm w}/c=1/\sqrt{40\, \dot{m}}$ \citep{2010MNRAS.402.1516K}. 
Since $\beta_{\rm w}\ll1$ for $\dot{m}\gg1$, the Lorentz factor of the wind is $\gamma_{\rm w}\approx1$; hence, there is no dependence of the wind properties with the line of sight ($\Theta$) due to relativistic effects, so $1-\beta_{\rm w} \cos \Theta \approx1$ \citep[see e.g.][]{Abaroa&Romero_RevMex_2024}. However, we note that there could be a dependence on the viewing angle because of the narrow conical low-density funnel \cite[with semi-opening angles $\lesssim 15^{\circ}$ in some cases, as recently shown by][for Cygnus-X3]{Veledina_etal_cygx3_2024NatAs} that forms due to the magneto-centrifugal barrier around the z-axis above the black hole and causes the geometric beaming \citep{2001ApJ...552L.109K,2009MNRAS.393L..41K,2010MNRAS.402.1516K,Kaaret_2012_ulx,Fabrika_etal_2012_ULX,Abaroa_etal_2023}. A low tilt with respect to the line of sight would allow the hard X-ray thermal emission to escape from the inner accretion disk and reach the observer, but this is not the case for S26 where a high inclination is inferred from radio and X-ray observations \citep{2010Natur.466..209P, soria2010}.

The density of this wind at a distance $R$ from the BH is
\begin{equation} \label{eq: densidad_viento}
    \rho_{\rm w}(r)=\dot{M}_{\rm w}/4\pi R^2 v_{\rm w}.
\end{equation}
where $R=\sqrt{r^2+z^2}$, with $r$ and $z$ cylindrical coordinates. We assume a steady state so that $v_{\rm w}$ and $\dot{M}_{\rm w}$ are constants and therefore the density depends only on the distance to the BH.



The wind itself is opaque until it reaches the photospheric radius, where it becomes transparent to its own radiation \citep{Poutanen_2007,2009PASJ...61.1305F,2015PASJ...67..111T,Zhou_etal_2019}. The apparent photosphere is defined as the surface where the optical depth $\tau_{\rm photo}$ is unity for an observer at infinity. Its height $z_{\rm photo}$, measured from the equatorial plane, can be found by integrating over the wind density \citep{2009PASJ...61.1305F}: 
\begin{equation}
    \tau_{\rm photo}=\int^\infty_{z_{\rm photo}} \gamma_{\rm w}(1-\beta_{\rm w} \cos{\Theta}) \, \kappa_{\rm co} \,\rho_{\rm co} {\rm d}z \approx \int^\infty_{z_{\rm photo}}  \kappa_{\rm co} \,\rho_{\rm co} {\rm d}z =1,
\end{equation}
where $\kappa_{\rm co}\approx \kappa$ is the opacity and $\rho_{\rm co}\approx \rho_{\rm w}$ (the sub-index 'co' refers to the comoving frame). As we assume a fully ionized wind, the opacity is dominated by free electron scattering ($\kappa=\sigma_{\rm T}/m_{\rm p}$, with $\sigma_{\rm T}$ the Thomson scattering and $m_{\rm p}$ the proton mass). We note that the location of the wind photosphere is completely determined by the rate of mass loss, and thus by the accretion input of matter.

We assume an accretion rate input of $\dot{M}_{\rm input}=10\,\dot{M}_{\rm Edd}$, which leads to a photosphere with a height of $z_{\rm photo}\approx 2.6\times 10^{9}\,{\rm cm}$ above the black hole. \hyperref[fig: photosphere]{Fig.~\ref{fig: photosphere}} shows the geometry of the photosphere in Cartesian coordinates, in units of the gravitational radius. The color map represents the temperature, which is $\sim 3.9\times 10^5\,$K at the top. In the present model for high inclination angles, the photosphere completely obscures the surroundings of the black hole, and thus the X-ray emission from the inner disk cannot reach the observer.\footnote{Note that the narrow, optically thin evacuation funnel, characteristic of ULXs, is not shown in Fig. \ref{fig: photosphere} because our source has a high inclination.} We estimate the temperature and emission from the wind photosphere in Sect. \ref{sec: wind emission}, along with all the other radiative processes.


\subsection{Jet} \label{sec: jet}


Two highly collimated jets are generated near the black hole in opposite directions, expanding laterally at the speed of sound.    The jets are perpendicular to the orbital plane of the binary system. In the following sections, we will discuss the characterization of jet power and launch, the magnetic field and matter density, as well as the acceleration of particles at the base and in the terminal regions of the jet.

\subsubsection{Jet: available power and launching}

Using general relativistic magnetohydrodynamics simulations, \cite{2018MNRAS.474L..81L} shows that, under certain physical conditions, a large-scale poloidal magnetic field can be generated from a purely toroidal field in the disk. The poloidal magnetic flux accumulates around the black hole until it becomes dynamically important, leading to a magnetically saturated -- also called ``arrested'' in the literature -- accretion disk \citep{2003PASJ...55L..69N}. This happens at a certain radius $r_{\rm d}\le r_{\rm ms}$. 


According to the disk-jet coupling hypothesis of  
\cite{1995A&A...293..665F}, the kinetic power of each jet is assumed to be proportional to the accretion power at the magnetically saturated radius \citep[where the poloidal magnetic flux becomes dynamically important, see][]{Begelman_etal_2022}, $L_{\rm j}\sim q_{\rm j}\,L_{\rm acc}(r_{\rm ms})$, where $q_{\rm j}$ is the efficiency in transferring power from the disk to the jet. We assume that the disk is saturated at a distance from the black hole of $r_{\rm ms}=100\,r_{\rm g}$. At this radius, a BH of $10M_{\odot}$ accretes matter at $\dot{M}_{\rm acc}(r_{\rm ms})=2.5\dot{M}_{\rm Edd}=3. 5\times10^{19}\,{\rm g\, s^{-1}}$, so the accretion power is $L_{\rm acc}(r_{\rm ms})=3.15\times10^{40}\,{\rm erg\, s^{-1}}$. This luminosity is not sufficient to explain the jet power in S26: some energy should also be extracted from the ergosphere of the rotating black hole, which spins down as the rotational energy is transferred to the jet \citep{1977MNRAS.179..433B}. In such cases, the efficiency of converting accretion power into jet power can exceed 100 percent in highly rotating black holes, reaching transient values of 300 percent in the most extreme cases \citep{2010ApJ...711...50T,2011MNRAS.418L..79T,2012MNRAS.423.3083M}. Considering that the jet kinetic luminosity of S26 estimated by \cite{2010Natur.466..209P} is $L_{\rm j}=5\times10^{40}\,{\rm erg\, s^{-1}}$, then the jet-disk coupling constant should be $q_{\rm j}=L_{\rm j}/L_{\rm acc}\approx 1.60$.

We assume that the jet is fully formed at a distance $z_0=100\,r_{\rm g}$ from the BH and with a radius $r_0=0.1\,z_0=10\,r_{\rm g}$. Afterward, the outflow expands as a cone of radius $r_{\rm j}(z)=r_0(z/z_0)=0.1\,z$ (that is, with a typical semi-opening angle of $\theta_{\rm j}\approx 5.7^{\circ}$). We adopt a moderate value for the Lorentz factor $\gamma_{\rm j}=(1-v_{\rm j}^2/c^2)^{-1/2}=3$, so $v_{\rm j}=0.94\,c$.

Although the jet starts as a Poynting flux outflow, a fraction $q_{\rm rel}\approx0.01$ of the jet power is expected to be in the form of relativistic particles at distances greater than $\sim$1000 $r_{\rm g}$ \citep[see][for the matter loading mechanisms]{2020Univ....6...99R}. Then we can write $L_{\rm rel}=q_{\rm rel}\,L_{\rm j}$. We include both the hadronic and leptonic fractions, $L_{\rm rel}=L_{\rm p}+L_{\rm e}$. The energy distribution between hadrons and leptons is unknown; we assume an equipartition near the base of the jet ($L_{\rm p}=\,L_{\rm e}$), and a hadronic-dominant scenario in the terminal regions ($L_{\rm p}=100\,L_{\rm e}$), since strong matter entrainment along the jet path is expected at distances of $\sim 175$ pc from the BH \citep[e.g.,][]{2005A&A...432..609B,2009A&A...497..325B}. At such distances the lobes formed by the jet are exposed to the interstellar medium with conditions similar to those found in supernova remnants, where the hadron-to-lepton ratio is 100, in agreement with rough estimates of cosmic ray composition determined in the vicinity of the Earth \citep[e.g.,][]{Beresinsky1969}.

\subsubsection{Jet: magnetic field and matter density}

The magnetic field $B_0$ near the BH can be estimated from the Blandford-Znajek power $P_{\rm BZ}$ of the jet  \citep{1977MNRAS.179..433B,2020Univ....6...99R}:
\begin{equation}\label{eq: pbz}
    P_{\rm BZ} \sim 10^{46}\,\left(\frac{B_0}{10^4\,{\rm G}}\right)^2\,\left(\frac{M_{\rm BH}}{10^9\,{\rm M_{\odot}}}\right)^2\, a^2 \,{\rm erg\,s^{-1}},
\end{equation}
where $0\le a<1$ is the normalized spin of the BH. This is a 1st-order approximation, valid for low spin values: if $a>0.5$, then the above equation underestimates the jet power by a factor of $\sim3$. As we are dealing with a highly rotating BH to account for the observed jet power in S26, we apply this factor $(L_{\rm j}\sim3\,P_{\rm BZ})$ to correct the deviation from the more accurate 6th-order formula derived by \cite{2010ApJ...711...50T,2011MNRAS.418L..79T}  \citep[see also Chapter 3 of][for details]{2015contopoulos}. We obtain a magnetic field of $B_0\approx 1.61\times 10^9\,{\rm G}$ near the black hole.

The magnetic field in the jet decreases with distance $z$ from the BH: if the flow expands adiabatically,
\begin{equation} \label{eq: magnetic field}
    B(z)=B_0\left(\frac{z'}{z}\right)^m,
\end{equation}
where $z'\approx 10\,r_{\rm g}$ and $1\le m\le 2$ is the magnetic index, which depends on the topology of $B$. We assume a mixed topology of the magnetic field -- poloidal and toroidal -- beyond $z'$, so $m=1.5$. The magnetic energy density is given by
\begin{equation}\label{eq: magnetic energy density}
    e_{\rm m}(z)=\frac{B^2(z)}{8\pi}.
\end{equation}
The force exerted by the toroidal field helps to collimate the jet, where collimation is the degree of parallelism of the streamlines. A cylindrical jet with all streamlines parallel to the symmetry axis is then perfectly collimated \citep{Romero-Vila2014}. Note that collimation is different from confinement, which refers to the width of the outflow\footnote{See \cite{2021APh...12802557C} for a description and schematic representation of the different zones in jets.}.

The number density of cold protons at a distance $z$ from the BH in the jet is \cite[e.g.][]{2019A&A...629A..76S}
\begin{equation} \label{eq: number density}
    n_{\rm p}(z)\sim \frac{\dot{M}_{\rm j}}{\pi\, r^2_{\rm j}(z)\, m_{\rm p}\,v_{\rm j}},
\end{equation}
where the mass loss rate in the jet is given by $\dot{M}_{\rm j}=L_{\rm j}/(\gamma_{\rm j}-1)c^2$. The energy density of matter reads 
\begin{equation}\label{eq: matter energy density}
    e_{\rm p}(z)=\frac{\dot{M}_{\rm j}}{2\pi z^2} v_{\rm j}.
\end{equation}

\subsubsection{Jet: shocks}

The X-ray and radio maps of S26 show nonthermal emission from the core and the lobes \citep{2010Natur.466..209P,soria2010}. Therefore, we consider two acceleration regions for relativistic particles in the jet: one at the base of the jet, but above the photosphere of the disk-driven wind, and the other at the terminal region of the jet. The acceleration is assumed to be due to diffusive shock acceleration in the matter-dominated regions of the system. We neglect further shocks or other acceleration mechanisms (such as magnetic reconnection) that may occur, since the emission regions are located in shock-prone regions in the observations.

Internal shocks can be produced near the base of the jet \citep{Kaiseretal2000A&A}. These shocks occur when the jet is matter dominated, i.e. when the energy density of the matter is greater than the magnetic energy density, $e_{\rm p}(z)>e_{\rm m}(z)$. By comparing Eqs. \ref{eq: magnetic energy density} and \ref{eq: matter energy density}, we find that this happens at a height $z_{\rm b}\sim 5\times 10^9\,{\rm cm}$, where we assume that particle acceleration occurs. Note that this shock region is located above the wind photosphere derived in the previous section ($z_{\rm b}> z_{\rm photo}$).
 

At the head of the jet, two more shocks are generated: a forward radiative shock that propagates into the interstellar medium (ISM), and a reverse adiabatic shock that propagates into the jet. Particles escaping from the reverse shock form the thermal cocoon of the source. Particle acceleration to relativistic energies occurs only in the reverse adiabatic shock. 
Conversely, in the radiative shocks, the gas emits large amounts of thermal, optically thin radiation, and no particle acceleration is produced. The calculations of the radiation and its partial absorption are detailed in the next section.

\section{Radiative processes} \label{sect: radiative processes}


In the following, we characterize the different radiative processes that we have calculated for S26. First, we describe the thermal emission (from the disk-driven wind and the bow shocks of the jets), and then the nonthermal radiation (inner jet and reverse shocks in the terminal region). The processes studied include the acceleration, escape, cooling, distribution, and radiation of different types of particles. At the end of the section, the absorption of the radiation in the source is described.

\subsection{Wind emission} \label{sec: wind emission}

We calculate the thermal emission from the photosphere of the disk-driven wind assuming that it expands at a constant rate equal to its terminal velocity. Since the mass loss rate of the disk is much higher than the critical rate, the wind is optically thick, and the emission from the disk is thermalized and re-emitted as a blackbody spectrum at its photosphere \citep{Zhou_etal_2019}. Note, however, that in the central region, as we already have mentioned, the dense wind surrounds a low-density funnel, where the hard X-rays from the central accretion disk can escape to reach a face-on observer \textbf{(see Sect. \ref{sec: wind photosphere})} \citep{2010MNRAS.402.1516K,Zhou_etal_2019}. 

The photosphere temperature measured by an observer at infinity is given by \citep{2009PASJ...61.1305F}:
\begin{equation}
     \sigma_{\rm T} T_{\rm w}^4=\frac{\dot{e} \, L_{\rm Edd}}{(1-\beta \cos{\Theta})^4 \, 4 \pi R^2} \approx \frac{\dot{e} \, L_{\rm Edd}}{4 \pi R^2},
\end{equation}
where $R=\sqrt{r^2+z^2}$. The radiative power of the wind in supercritical sources is limited to the Eddington luminosity \citep{Zhou_etal_2019}, so we assume that the comoving luminosity of the wind is a fraction $\dot{e}=0.1$ of $L_{\rm Edd}$ \citep{2022A&A...664A.178S}. The temperature of the wind at the photospheric height is then $T_{\rm photo}=3.9\times10^{5}\,{\rm K}$.

\subsection{Thermal lobes}

The radiative forward shocks generated at the jets' terminal regions, as they interact with the ISM, will produce thermal extended emission due to the free-free interaction of the particles (thermal Bremsstrahlung). The emissivity of a thermal, optically thin plasma at a temperature $T$ for a frequency $\nu$ is given by \citep{2011hea..book.....L}:
\begin{equation}
    \kappa_{\nu}= \mathbb{C}\,Z^2\, T^{-1/2}\,N_{\rm i}\, N_{\rm e}\, 
    g(\nu,T)\, \exp{\left(\frac{-h\nu}{k_{\rm B}T}\right)}\ {\rm erg\,s^{-1}\,cm^{-3}\,Hz^{-1}},
\end{equation}
where $\mathbb{C}=6.8\times10^{-38}$, $Z$ is the atomic number, $N_{\rm i}$ and $N_{\rm e}$ are the number density of ions and electrons (in units of $\rm cm^{-3}$), $h$ is the Planck constant, and $k_{\rm B}$ the Boltzmann constant. The Gaunt factor $g(\nu,T)$, for X-ray energies, is \citep{2011hea..book.....L}:
\begin{equation}
    g(\nu,T)\approx
    \left\lbrace \begin{array}{l}
  \dfrac{\sqrt{3}}{\pi}\, \ln{\left(\dfrac{k_{\rm B}T}{h\nu}\right)}, \ \ h\nu/k_{\rm B}T \ll 1\\ \\
 \sqrt{\frac{h\nu}{k_{\rm B}T}}, \ \ h\nu/k_{\rm B}T \gg 1
    \end{array} 
    \right.
\end{equation}
The luminosity of a region of volume $V$ as a function of the frequency is then $L(\nu)=V\,\kappa_{\nu}\, \nu$. This thermal radiation from the lobes of S26 is identified with the observed soft X-ray photons detected by the \textit{Chandra} and \textit{XMM-Newton} telescopes.

\subsection{Acceleration and escape timescales}

The diffusive acceleration rate of the particles is given by \citep[e.g.,][]{1999tcra.conf..247P}:
\begin{equation}
    t^{-1}_{\rm{acc}}=\eta_{\rm acc}~\frac{e\,Z\,c\,B}{E}
\end{equation}
where $e$ is the electric charge, $Z$ the atomic number, $c$ the speed of light, $B$ the magnetic field, and $E$ is the energy of the particle. The acceleration efficiency $\eta_{\rm acc}$ of the process can be estimated with the velocity of the jet, assuming Bohm diffusion: $\eta_{\rm acc}=3\beta_{\rm sh}^2/8$, where $\beta_{\rm sh}=v_{\rm sh}/c$, with $v_{\rm sh}$ the shock velocity in the corresponding region: in the case of the base, $v_{\rm sh}\equiv v_{\rm j}$ while in the reverse shock $v_{\rm sh}\equiv v_{\rm rs}$ (see \hyperref[tab: parametros generales]{Table~\ref{tab: parametros generales}}). These values lead to $\eta_{\rm acc}\approx0.1$. 

Escape is advective at the base of the jet (particles are removed from the acceleration region by the bulk motion of the jet) and diffusive in the lobes:
\begin{equation}
\begin{array}{l}
   t_{\rm adv}^{-1}=\left(\dfrac{\Delta x_{\rm acc}}{v_{\rm j}}\right)^{-1} \ \ {\rm (advection)}\\ \\
 t_{\rm diff}^{-1}=\left(\dfrac{\Delta x_{\rm acc}^2}{D(E)}\right)^{-1}  \ \ {\rm (diffusion)}
    \end{array} 
\end{equation}
where $\Delta x_{\rm acc}$ is the size of the corresponding acceleration region and $D(E)$ is the diffusion coefficient, considered in the Bohm regime, so 
\begin{equation}
    D(E)=D_{\rm B}(E) =\frac{E\,c}{3\,B\,e}.
\end{equation}

Accelerated particles can cool by several different processes, which are described in the next subsection. The maximum energy for each type of particle can be deduced by looking at the point where the acceleration rate equals the total cooling or escape rate. This energy cannot exceed the maximum energy locally imposed by the Hillas criterion, $E^{\rm max}_{\rm e,p}<E^{\rm max}_{\rm Hillas}=e\,Z\,\Delta x_{\rm acc}\,B$.  

\subsection{Energy losses}

The timescales associated with cooling are related to the total energy loss of the particles:
\begin{equation}
    \frac{dE}{dt}\approx \frac{-E}{t_{\rm{cool}}},
\end{equation}
where the total cooling rate is
\begin{equation}
    t_{\rm{cool}}^{-1} = \sum_i t_{i}^{-1},
\end{equation}
where $t_i$ is each timescale of the cooling processes involved.


Adiabatic losses may also be relevant at the internal shock, because of the pressure exerted by the particles to expand the lateral walls of the jet,
\begin{equation}
    t_{\rm ad}^{-1}=\left(\frac{3\Delta x_{\rm acc}}{2v_{\rm j}}\right)^{-1}.
\end{equation}


Radiative cooling is caused by various nonthermal processes as a consequence of the interaction of particles with radiation, magnetic field, and matter. The intensity of the magnetic field in the shocks at the base of the jet is calculated according to Eq. \ref{eq: magnetic field}; at the reverse shock it is fixed to a typical value of the ISM as $2.5\,{\mu}$G. In addition, we consider an enhancement of $\sim 4$ due to compression \citep{1999isw..book.....L}. The dominant photon fields are provided by the photosphere of the disk-driven wind (for the base of the jet), which decays with $1/d^2$, and the cosmic microwave background (for the terminal region). The cold matter density is calculated according to Eq. \ref{eq: number density} for the base of the jet, and is set to $0.1\,{\rm cm^{-3}}$ for the terminal region \citep{soria2010}.

Our model is lepto-hadronic, and so we calculate the following cooling processes  numerically:

--Synchrotron: interaction of protons and electrons with the ambient magnetic field.

--Inverse Compton (IC): collision of relativistic electrons with photons of the ambient radiation field.

--Relativistic Bremsstrahlung: Coulombian interactions between relativistic electrons and cold matter.

--Photo-hadronic interactions: Interaction of highly relativistic protons with photons of the ambient radiation field.  This interaction produces pions, which decay into gamma rays, $e^+e^-$ pairs, and neutrinos.

--Proton-proton: Collision of relativistic protons with cold matter. This process is also a source of gamma rays, pairs, and neutrinos.

\subsection{Particle distributions} \label{subsec: particle distribution}

The relativistic particles have a distribution given by ${\rm d}N= n(\vec{r},E,t){\rm d}E{\rm d}V$, where $n$ is the number density of particles, $t$ the time, $\vec{r}$ the position, $V$ the volume, and $E$ the energy. The evolution of this distribution is determined by the transport equation \citep[see][]{1964ocr..book.....G}. We solve this equation numerically in steady state and in the one-zone approximation: 
\begin{equation}
    \frac{\partial}{\partial E}\left[\frac{{\rm d}E}{{\rm d}t} N(E)\right]+\frac{N(E)}{t_{\rm esc}}=Q(E),
\end{equation}
where $t_{\rm esc}$ is the corresponding escape time, 
and the particle injection function,
\begin{equation} \label{eq: injection} 
    Q(E)=Q_{0}E^{-p}\exp{(-E/E_{\rm max})},
\end{equation} 
is a power-law in the energy with an exponential cutoff and a spectral index of $p\sim 2$, which is characteristic of the Fermi first-order acceleration mechanism \citep[see e.g.,][]{1983RPPh...46..973D}. The normalization constant $Q_0$ is obtained from the available power for the different particle species
\begin{equation}
    L_{\rm rel\,(\rm e,p)}=V_{\rm acc} \int^{E^{\rm max}_{\rm (e,p)}}_{E^{\rm min}_{\rm (e,p)}}E_{\rm (e,p)}\,Q_{\rm (e,p)}(E_{\rm (e,p)})\,{\rm d}E_{\rm (e,p)},
\end{equation}
where $V_{\rm acc}$ is the volume of the acceleration region, and $E^{\rm max}_{\rm (e,p)}$ the maximum energy reached by protons and electrons.

\subsection{Nonthermal emission}

Once we have the particle distributions, we calculate the spectral energy distribution (SED) for each of the relevant processes involved in the cooling. This radiation corresponds in particular to the nonthermal emission of S26 detected from the core (X-rays) and the lobes (radio emission). The reader is referred to \cite{2008A&A...485..623R}, \cite{Romero&Paredes2011}, \cite{Romero-Vila2014}, and references therein for additional details on the radiative processes.

Relativistic Doppler affects the emission produced at the internal shock due to the relativistic motion of the gas. The Doppler factor depends on the angle $i$ of the jets with respect to the line of sight:
\begin{equation}\label{eq: doppler}
    D=\frac{1}{\gamma_{\rm j}(1\pm \beta_{\rm j}\cos{i})},
\end{equation}
where the sign $\pm$ accounts for the jet $(-)$ and the counterjet $(+)$. The apparent luminosity is then $D^2$ times the comoving luminosity,
\begin{equation}
    L_{\rm app} = D^2\,L_{\rm emitted}.
\end{equation}
We get $D^2\approx0.19$ for the jet and $\approx 0.07$ for the counterjet (cj), so the apparent luminosity is less than the emitted. Since $L_{\rm cj}\sim 0.4\,L_{\rm j}$, then $L_{\rm j+cj}\sim 1.4\, L_{\rm j}$. Note also that since the lobes are static, their emission is not affected by relativistic effects $(D\equiv 1)$.

\subsection{Absorption} \label{subsec: absorption}

Finally, we calculate the absorption of gamma rays by pair creation from photon-photon annihilation, $\gamma + \gamma \rightarrow e^+ + e^-$. The nonthermal photons on their way out of the emission region can find thermal ambient photons and annihilate. High-energy radiation emitted in the jets is also attenuated by $\gamma \gamma$--annihilation with synchrotron photons produced internally. The absorption is quantified by the optical depth of the medium, $\tau_{\gamma \gamma}$, which depends on the energy. 
The attenuated, observed, luminosity reads:
\begin{equation}
    L_{\rm obs}(E_\gamma)=L_{\rm app}(E_\gamma)\cdot {\rm e}^{- \tau},
\end{equation}
where ${\rm{e}}^{-\tau}$ is the attenuation factor.
We refer to Sect. 4.5 of \cite{Abaroa_etal_2023} for details in the calculation of the absorption.

\section{Results} \label{sect: results}
We present here the results of applying our model to determine the thermal and nonthermal emission of S26, from the core and the lobes. 

\subsection{Energy gain and losses}
\begin{figure*}[ht]
\begin{multicols}{2}
\centering
    \includegraphics[width=8cm]{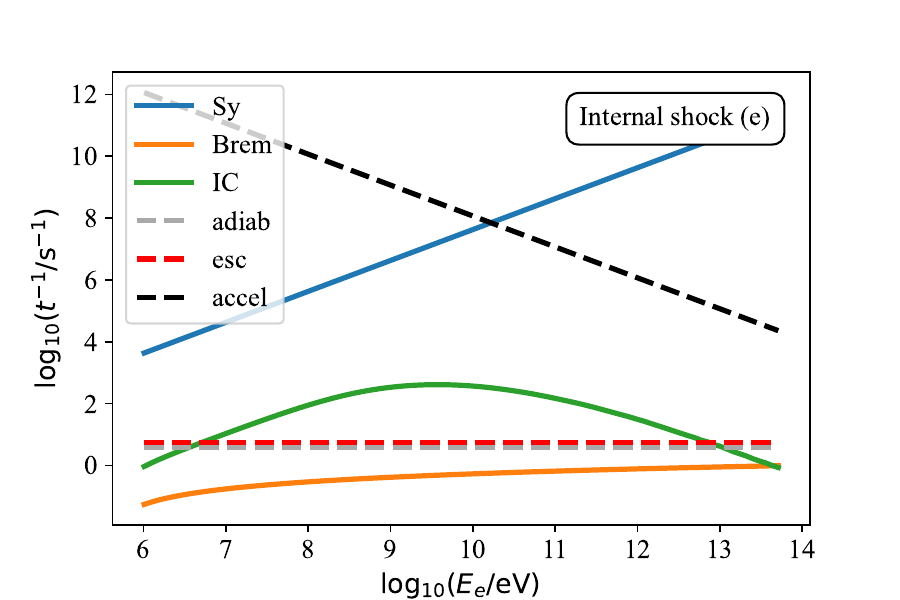}\par 
    \includegraphics[width=8cm]{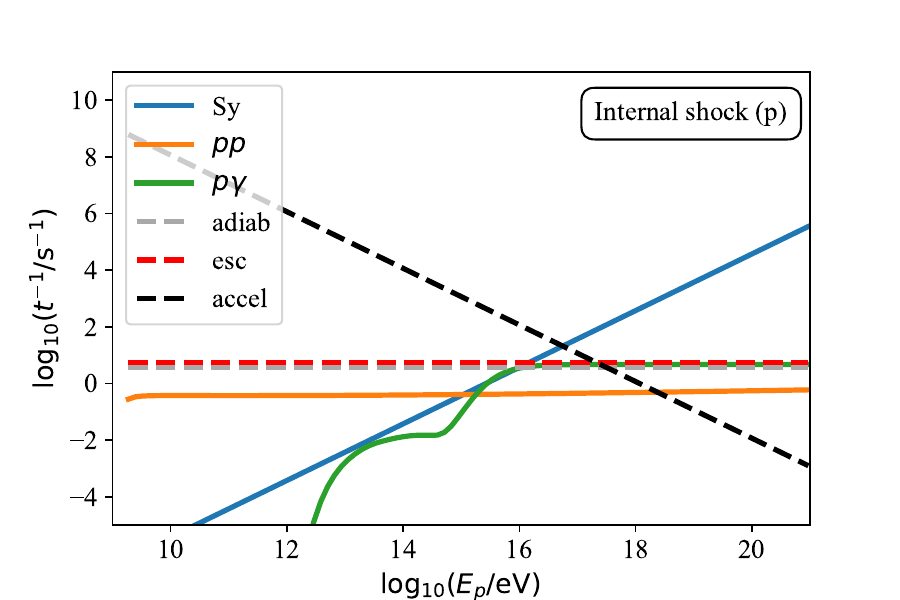}\par 
     \includegraphics[width=8cm]{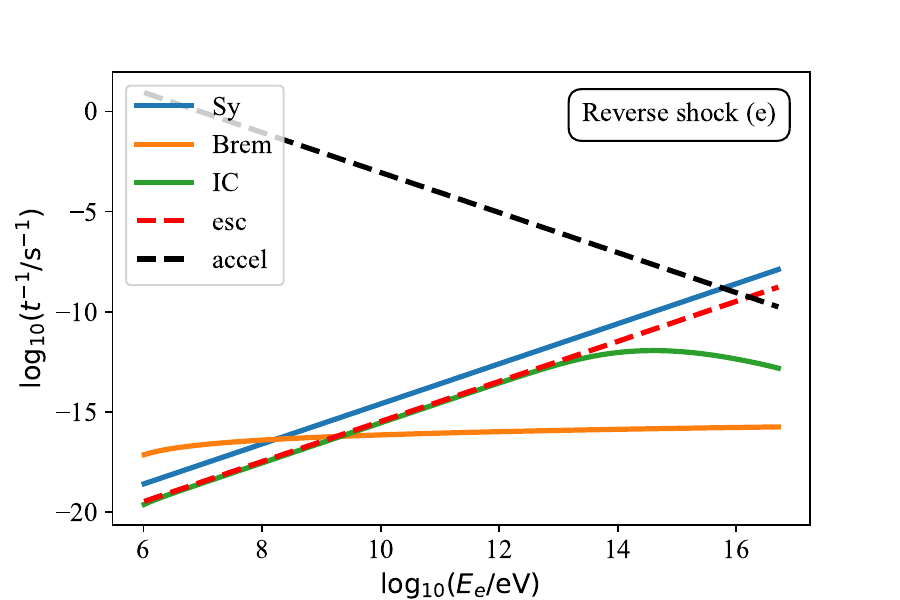}\par 
    \includegraphics[width=8cm]{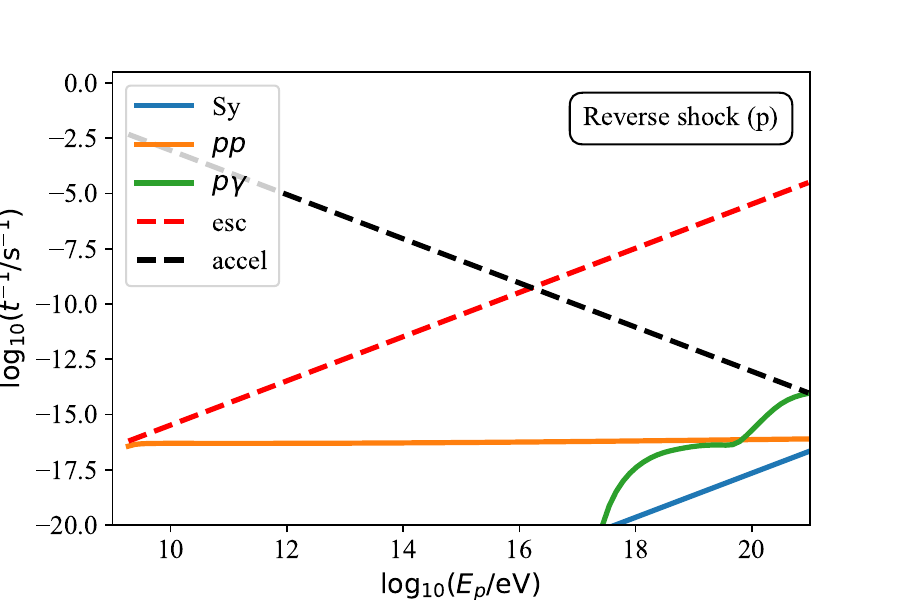}\par 
    \end{multicols}
\caption{Cooling timescales of particles accelerated at the base of the jet (left) and at the terminal region (right). We show the energy gains and losses for electrons in the top panel and the same for protons in the bottom panel. In the internal shock at the base of the jet, the particles cool mainly by the synchrotron mechanism due to the strong magnetic field, reaching energies of 10 GeV (electrons) and 50 PeV (protons). At the reverse shock in the terminal region of the jet, electrons also cool mainly by synchrotron radiation, reaching energies of 5 PeV, while protons reach energies of $\sim 10\,$PeV. The efficiency of the acceleration in all cases is $\eta_{\rm acc}\sim 0.1$.}
\label{fig: cooling}
\end{figure*}

\hyperref[fig: cooling]{Figure~\ref{fig: cooling}} shows the timescales associated with particle acceleration, escape, and cooling. We have calculated these processes for the base of the jet (left) and the reverse shocks in the lobes (right), in each case for both leptons (top) and hadrons (bottom). The efficiency of the acceleration in all cases is $\eta_{\rm acc}\sim 0.1$.

In the internal shock at the base of the jet, electrons cool mainly by synchrotron (as expected, because of the strong magnetic field), reaching a maximum energy of $\sim 10\,$GeV, whereas protons reach a maximum energy of $\sim  50\,$PeV, being synchrotron and advective escape the dominant timescales for them. The Hillas maximum allowed energy is $E_{\rm Hillas}^{\rm max}=2\times 10^{17}\,{\rm eV}$ in this region. 

Electrons accelerated in the reverse shock in the lobes also cool mainly by synchrotron radiation, reaching a maximum energy of $\sim 5\,$PeV. Protons are accelerated to energies of $\sim 10\,$PeV, and their dominant loss is diffusive escape. 
These ultra-relativistic protons, once they have escaped, will in turn propagate through the interstellar medium and will contribute to the cosmic ray (CR) population in the host galaxy of the MQ.

\subsection{Nonthermal radiation}

\begin{figure*}[ht]
\begin{multicols}{2}
    \centering    \includegraphics[width=\columnwidth]{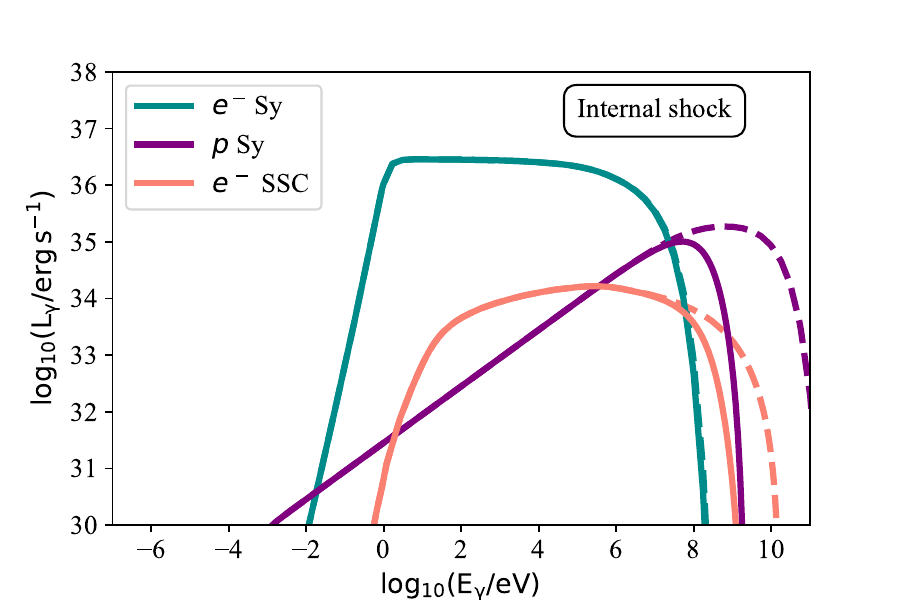}\par
    \includegraphics[width=\columnwidth]{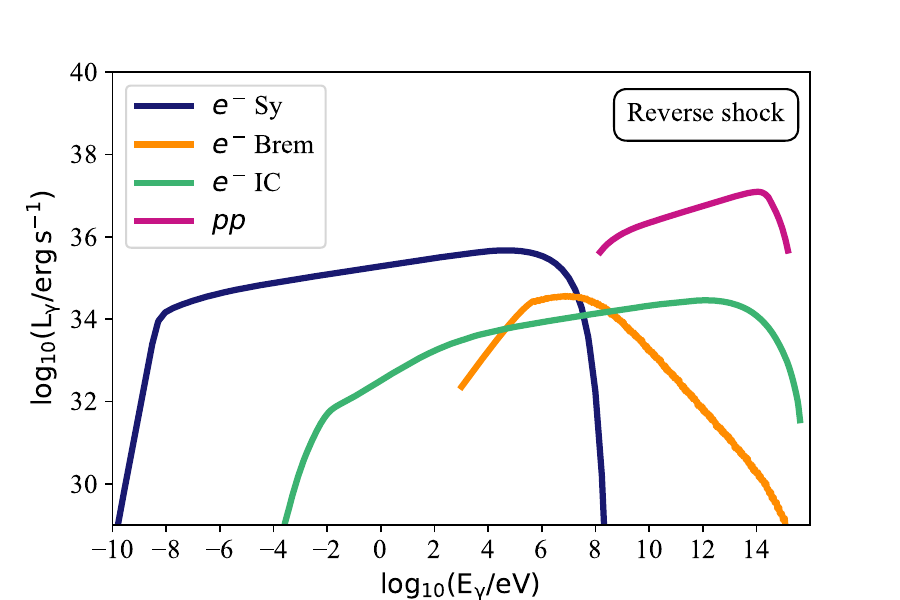}\par
    \end{multicols}
        \caption{Left: Nonthermal SEDs of radiation produced at the internal shock in the jet, close to the BH but above the disk-driven wind photosphere. The dominant mechanism is electron synchrotron. 
        For energies $>1\,$GeV the emission is completely attenuated due to internal absorption by pair creation. Dashed lines represent the unabsorbed emission. Right: Nonthermal SEDs of the radiation produced at the reverse shock in the terminal region of the jet. Electron-synchrotron radiation is the dominant process at low energies while proton-proton dominates at high and very high energies. No attenuation affects the radiation produced at the lobes.}
        \label{fig: individual_seds}

\end{figure*}

\hyperref[fig: individual_seds]{Figure~\ref{fig: individual_seds}} shows the nonthermal SEDs of the jet base (left) and the lobes (right). In the case of the base, the electron-synchrotron emission extends from $10^{-1}$ to $\sim 10^6\,$eV with an almost constant luminosity of $5\times10^{36}\,{\rm erg\, s^{-1}}$, while the proton-synchrotron mechanism provides emission up to $\sim 1\,$TeV, although internal photon-photon annihilation produces absorption for energies higher than $\sim 1\,$GeV. The beaming effects are already included. The electron-synchrotron mechanism is then responsible for the X-ray emission produced at the core, detected by \textit{Chandra} and \textit{XMM-Newton}. For the lobes, synchrotron, Bremsstrahlung, inverse Compton, and proton-proton interactions are all relevant. The luminosity produced by synchrotron is above $10^{34}\,{\rm erg\, s^{-1}}$ from very low ($10^{-9}\,$eV) to high ($10^{7}\,$eV) energies, while the emission due to $\pi^0$ decays through the $pp$ channel is above $10^{35}\,{\rm erg\, s^{-1}}$ in the $10^{8}-10^{14}\,$eV energy range. The electron-synchrotron mechanism is responsible for the radio emission in the hotspots detected with ATCA by \citet{soria2010}, and $pp$ and the subsequent decays are responsible for the high-energy gamma-ray emission.


\subsection{Total SED}

\begin{figure}[ht]
    \centering    \includegraphics[width=\columnwidth]{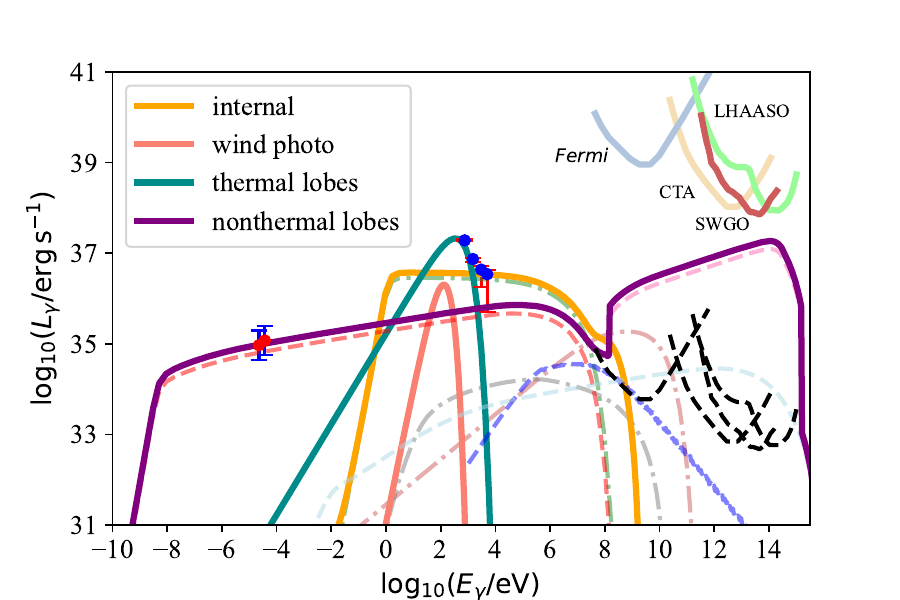}
    \caption{Luminosity of S26 as a function of the energy for our physical model with the parameters listed in Table \ref{tab: parametros generales}. We plot the thermal and nonthermal SEDs of the system. The thermal contributions are the photosphere of the disk-driven wind and the optically thin emission in the lobes. Nonthermal emission is produced in the internal shocks at the base of the jet (orange solid line, jet $+$ counterjet) and in the reverse shock of the lobes (purple solid line, both lobes). Individual nonthermal processes are shown with dashed and dotted-dashed lines (for details see \hyperref[fig: individual_seds]{Fig.~\ref{fig: individual_seds}}). We plot radio data from ATCA \citep[red dots,][]{soria2010} corresponding to the emission of the two lobes combined, and X-ray data from \textit{XMM-Newton} (blue dots, this work) corresponding to the core and lobes, for obsID 0748390901. Sensitivities of the \textit{Fermi}, CTA, SWGO and LHAASO gamma-ray detectors are also shown for two different distances: that from S26 (3.9 Mpc, solid colored lines), and a galactic distance of $\sim 10\,{\rm kpc}$ (black dashed lines).}
    \label{fig: total_sed}
\end{figure}

\hyperref[fig: total_sed]{Figure~\ref{fig: total_sed}} shows the total SED of S26, already corrected for absorption and in the observer's frame (on Earth). We can see the thermal bump caused by the wind photosphere and the forward shocks in the lobes. We also distinguish several plateaux due to the nonthermal processes (the total contribution of the internal shock in the jet and the backward shocks in the lobes). The emission from the photosphere of the disk-driven wind is calculated with an accretion input of $10\,\dot{M}_{\rm Edd}$. The optically thin emission from the radiative shocks in the terminal region of the jets fits the detected soft X-ray emission with a temperature of $\sim 4\times 10^6\,$K, while the nonthermal emission from the inner jet (plus the small contribution from the counterjet) fits well the hard X-ray emission \citep[\textit{Chandra} from][and \textit{XMM-Newton} from this paper]{2010Natur.466..209P}. The combined nonthermal radiation from both lobes provides a good fit to the radio emission detected with ATCA \citep{2010Natur.466..209P,soria2010}. The gamma rays from the lobes reach energies of $\sim10^{14}\,$eV. As the plot shows, the source cannot be detected at gamma-ray energies with current instruments, so it should be studied with radio interferometric arrays in the southern hemisphere (such as MeerKAT -originally known as the Karoo Array Telescope- and the future SKA -Square Kilometre Array-) and with space-based X-ray observatories.  

We note, however, that a similar MQ in our Galaxy (i.e., a super-Eddington MQ) could be detected at very high and ultra-high energies with \textit{Fermi}, CTA (Cherenkov Telescope Array), SWGO (Southern Wide-field Gamma-ray Observatory), and LHAASO (Large High Altitude Air Shower Observation). In fact, this is the case of SS433, a weaker source that has already been detected in VHE (very-high-energy) gamma rays \citep{SS433-2018Natur.562...82A,SS433-2024Sci}.


\subsection{Exploration of the parameter space} \label{sec: parameter_space}

\begin{figure}[ht]
    \centering    \includegraphics[width=\columnwidth]{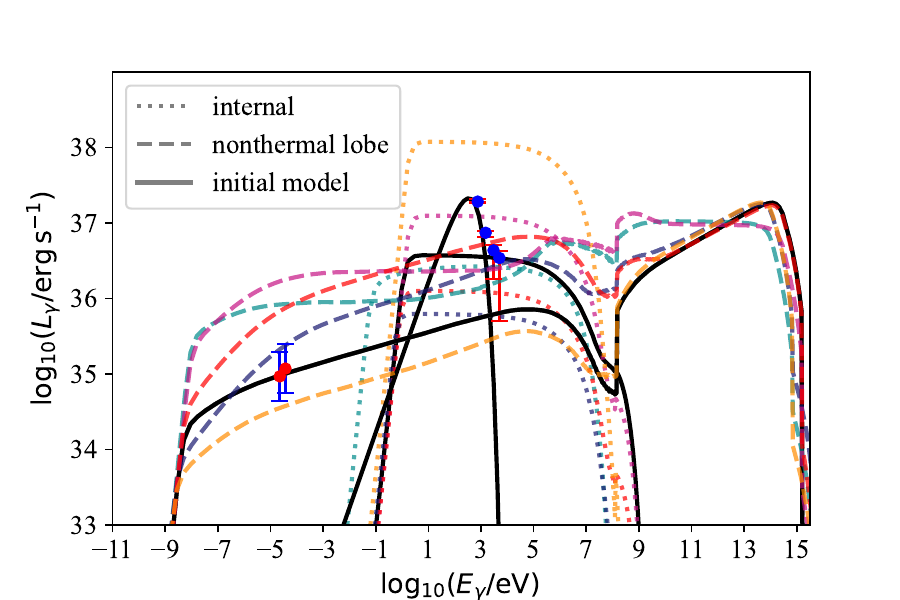}
    \caption{Exploration of the space parameter of the model. We plot SEDs for sets of input parameters different from those of Table \ref{tab: parametros generales} and Fig. \ref{fig: total_sed}. The dotted lines correspond to the SEDs of the different specific scenarios for the recollimation shock at the base $(b)$ of the jet, ($Sb1-Sb5$).  The dashed lines correspond to the SEDs of the scenarios of the reverse shock at the lobes $(l)$ of the jet, ($Sl1-Sl5$). The corresponding colors for both shocks are cyan, blue, pink, red, and orange for models 1 to 5, respectively. The different model parameterizations are described in Sect. \ref{sec: parameter_space}. Solid black lines represent our initial scenario. The observational data are plotted in the same way as in Fig. \ref{fig: total_sed}.}
    \label{fig: total_sed_alternative}
\end{figure}

We explore the parameter space of S26 to analyze the output spectra as we vary some of the most sensitive parameters. Fig. \ref{fig: total_sed_alternative} shows the SEDs we obtained for different scenarios, including the one shown in Fig. \ref{fig: total_sed} (in solid black lines) for comparison. We consider five different scenarios for the jet base ($Sb1-Sb5$, dotted lines) and the nonthermal lobe ($Sl1-Sl5$, dashed lines): 
\begin{itemize}    

\item $Sb1$ (cyan line): We only vary the location of the acceleration region, setting $z_{\rm b}=10^{11}\,{\rm cm}$, which results in a lower magnetic field 
but a more extended emission compared to our original scenario. The nonthermal X-ray emission is underestimated in this case. 

\item $Sb2$ (blue line): We vary the black hole mass $M_{\rm BH}=15 M_{\odot}$ and the hadron-to-lepton ratio, $K_{\rm b}=10$. The power results underestimated by almost an order of magnitude.


\item $Sb3$ (pink line): We vary the jet Lorentz factor $\gamma_{\rm j}=5$ and the fraction of energy going to relativistic particles $q_{\rm rel}=0.1$. This produce an overluminous source by a factor of $\sim5$.

\item $Sb4$ (red line): We vary the black hole mass $M_{\rm BH}=15 M_{\odot}$, the spin $a=0.9$, and the jet Lorentz factor $\gamma_{\rm j}=5$. The luminosity results strongly underestimated.

\item $Sb5$ (orange line):  We again vary the jet Lorentz factor $\gamma_{\rm j}=1.5$ and the fraction of the power going to relativistic particles $q_{\rm rel}=0.1$. The result is a powerful source that surpasses the actual observed one by nearly two orders of magnitude.

\item $Sl1$ (cyan line): We vary the magnetic field $B_{\rm l}=1\,\mu{\rm G}$, the spectral index $p_{\rm l}=2$, and the hadron-to-lepton ratio, $K_{\rm l}=10$. In this scenario the radio emission is overestimated by an order of magnitude.

\item $Sl2$ (blue line): We vary the magnetic field $B_{\rm l}=1\,\mu{\rm G}$, the number density $n_{\rm l}=1\,{\rm cm^{-3}}$, and the hadron-to-lepton ratio, $K_{\rm l}=10$. This scenario overestimates the radio emission by a factor of 2.

\item $Sl3$ (pink line): We vary the spectral index $p_{\rm l}=2$, the number density $n_{\rm l}=1\,{\rm cm^{-3}}$, and the hadron-to-lepton ratio, $K_{\rm l}=10$. This is the scenario that overestimates the overall emission the most.

\item $Sl4$ (red line): We vary the hadron-to-lepton ratio $K_{\rm l}=10$ and the number density $n_{\rm l}=1\,{\rm cm^{-3}}$. This scenario overestimates the emission by a factor of 2.

\item $Sl5$ (orange line): We vary only the magnetic field, $B_{\rm l}=1\,\mu{\rm G}$. The emission is underestimated by a factor of $\sim 3$.
\end{itemize}

As can be seen from Fig. \ref{fig: total_sed_alternative}, no alternative choice of parameter values at the base of the jet fits well with the nonthermal X-ray emission observed from the core of S26. The model shows sensitivity to the BH mass, the hadron-to-lepton ratio, the fraction of power going to relativistic particles, the acceleration region, or the Lorentz factor, all of which we vary to explore the parameter space. The luminosity of the radiation produced in this region varies uniformly over the energy range $\sim 10^{-1}-10^{6}\,{\rm eV}$ in all scenarios. The hadron-to-lepton ratio and the power going to the relativistic particles have a linear behavior with respect to the luminosity: a higher $q_{\rm rel}$ and a lower $K_{\rm b}$ leads to a higher synchrotron emission and vice versa. The jet Lorentz factor is one of the most sensitive parameters, since the luminosity in the observer's frame is corrected by Doppler effects: the higher $\gamma_{\rm j}$, the lower the emission in the direction of Earth, because of the high inclination of the source (see Eq. \ref{eq: doppler}). The location of the acceleration region, the mass of the black hole, and its spin have little effect on these alternative scenarios. 

In the case of the lobes, we varied the BH spin and mass, the number density, the magnetic field, and the hadron-to-lepton ratio. Again, the model is sensitive to changes in these parameters, showing in the most extreme case a difference of two orders of magnitude with respect to the observations. Furthermore, scenarios $Sl2$, $Sl3$, and $Sl4$ have nonthermal X-ray luminosities resolvable with \textit{Chandra}, but this telescope has not detected any nonthermal X-ray emission from the lobes. Unlike the emission from the base of the jet, the lobes do not show uniform variations across the energy spectrum. The curves vary at energies $>10^8$ eV, essentially due to a softening of the particle distribution (scenarios $Sl1$ and $Sl3$, where we adopt a spectral index of $p=2$), with the emission at $10^9$ eV being an order of magnitude higher than in the case of our initial scenario. The variation of the magnetic field and the hadron-to-lepton ratio does not lead to significant changes at these energies. On the contrary, the model is susceptible to these two parameters in the energy range $10^{-8}-10^8$ eV, where the synchrotron process dominates: a higher $B_{\rm l}$ and a lower $K_{\rm l}$ lead to a higher luminosity and vice versa. The black hole mass and its spin have little impact on these alternative scenarios of the lobes.

\section{Discussion} \label{sect: discussion}

Our model of S26, which assumes that this source is a super-Eddington accretor seen at a large angle of inclination, allows us to solve the puzzle posed by the large power inferred from the large scale jets and the paucity of X-rays from the central source. It is the dense wind ejected by the accretion disk that obscures most of the X-ray emission. Only nonthermal X-rays produced outside the photosphere of the wind can reach the observer. 

Future observations with different detectors could help to further constrain the free parameters of the models, such as the ratio of relativistic protons to electrons, the spectral index of the particles at different distances from the core, the magnetization of the jet, the black hole mass and spin, or the accretion rate. For example, the detection of radio emission from the core of the system could be used to better determine the location of the acceleration zone at the base of the jet and the strength of the magnetic field there. This would require interferometric observations with a higher resolution than the one that can be obtained with the ATCA. In addition, the detection of very high energy gamma rays from S26 could indicate a higher particle density at the end of the jets and place a lower limit on the energy of the CRs accelerated in the source. A flaring event caused by the interaction of the jet with overdensities of matter could also favor the detection at very high or ultra-high energies \citep{Araudo-et-al2009A&A,Owocki-et-al2009ApJ}.

As we mentioned in Sects. \ref{sec: intro} and \ref{sect: physical model}, we note that a source similar to S26, but with a different orientation with respect to the line of sight, would be seen as an ultraluminous X-ray source. The jet emission would also be magnified by a factor of $\gamma_{\rm j}^2$. This would increase the received flux from the inner jet by about an order of magnitude. For similar considerations about SS433 as a ULX when viewed from a different point of view, see \citet{Begelman-SS433-2006MNRAS}. 

The interaction of the strong wind ejected from the disk of S26 with nearby clouds can also lead to gamma-ray emission, as recently proposed for SS433 \citep{Li_etal_2020}. The equatorial outflow could propagate far beyond the system and be revealed if it collides with any clouds \citep[see][for a characterization of the wind]{Picchi_etal2020}. The shock generated by the collision would convert the kinetic energy of the plasmoids into internal energy and relativistic particles, which could then be cooled by $pp$ interactions with the cloud material. Such a scenario might explain the detection of a GeV source by the \textit{Fermi} satellite on the side of SS433 \citep{2020NatAs...4.1132B,Li_etal_2020}. Since S26 is at a distance of 3.9 Mpc, a luminosity $L_{\gamma}>10^{39}\,{\rm erg\,s^{-1}}$ is required to be detected by \textit{Fermi} at GeV energies. According to \cite{Li_etal_2020}, to account for this luminosity via hadronic interactions, we need a total energy of protons interacting with a nearby cloud $W_{\rm p}=2.5\times10^{48}(L_{\gamma}/10^{34}\,{\rm erg\,{s^{-1}}})(n/20\,{\rm cm^{-3}})\,{\rm erg}=2.5\times10^{53}\,{\rm erg}$, where $n$ is the number density of the cloud. The kinetic power of the wind is $L_{\rm w}=\dot{M}_{\rm w}v^2/2=1.4\times 10^{38}\,{\rm erg\,s^{-1}}$; therefore, the system should be injecting wind into the cloud for a time $t_{\rm inj}=W_{\rm p}/L_{\rm w}\sim 3\times 10^7 \,{\rm yr}$, which is much longer than the estimated age of the microquasar $(t=2\times 10^5\,{\rm yr})$, so the emission due to wind-cloud interactions would be undetectable with any reasonable choice for the values of the parameters in the model.  

Another interesting result of our models is that S26 can accelerate protons to PeV energies both in the inner jet and in the extended lobes. The possibility of powerful microquasars producing high-energy cosmic rays has been addressed in several papers \citep[see e.g.][]{Heinz&Sunyaev_2002,2005A&A...432..609B,Fender_etal_2005,Cooper_etal_2020}. The protons can escape from the jet through $p+p\rightarrow p+n+\pi^+$ or $p+\gamma\rightarrow n+\pi^+$ channels: the neutrons are no longer confined by the strong magnetic field and can decay far away from the sources, releasing very relativistic protons into the ISM, where they will diffuse until they interact with the surrounding matter and produce gamma rays \citep{Escobar2022A&A}. The source, however, remains mostly obscured by the wind of the accretion disk. This type of model could be relevant for the recently detected PeVatrons in our Galaxy. 

Current observations at radio wavelengths and X-rays provide us with information to constrain the parameters of our physical model, both for the core of the system and for the thermal and nonthermal lobes. Although we cannot in principle guarantee a direct relation between the observations and the acceleration of CRs to PeV energies, the results we have obtained for the emission produced by the system (which fits the observational data well), correspond to a set of parameters  (see Table \ref{tab:Tabla_model_parameters}) that lead to the acceleration of protons to ultra-high energies at the reverse shock in the lobes, where the size of the accelerator is one of the relevant quantities involved in the process.

The LHAASO catalog lists 43 galactic ultra-high energy (UHE, $E>0.1\,$PeV) gamma-ray sources \citep{2024ApJS..271...25C}. Observations of UHE gamma rays provide a key clue to the acceleration of PeV protons and thus to the origin of galactic CRs. The detection of gamma rays above 10-100 TeV is crucial for tracing the accelerators \citep{2016JPhCS.718e2043V,2022icrc.confE..11C} and determining the highest energies of CRs in the Galaxy. 

We propose that yet-undetected super-Eddington MQs could be some of these sources (whose exact nature remains unknown). In the case of S26, the power of the system may allow protons to reach energies greater than 1 PeV, as we have shown. A key feature of such powerful jets relies on the final size of the accelerator in the reverse shock at the lobes. In our model, the maximum energy of the relativistic protons from the terminal region is limited by the diffusive escape (Fig. \ref{fig: cooling}, bottom right), which depends on the square of the size of the acceleration region. A powerful jet could allow the acceleration region to be larger, imposing a cutoff to the protons at quite high energies. However, this scenario is conditioned by the physics in the system, which is quantified by several not well-known parameters, such as the efficiency of the acceleration processes in the different environments and the strength of the magnetic field in the acceleration regions. Nevertheless, our results suggest that super-Eddington MQs could potentially serve as PeVatron sources as long as they maintain their super-Eddington regime \citep{2024arXiv240106271B}.

Furthermore, if a Galactic super-Eddington microquasar is located in a star-forming region, it could very easily go undetected if its jets are misaligned with the line of sight. The central X-ray emission would be obscured by the presence of the opaque wind, and proton-dominated jets tend to be dark most of the way \citep[see e.g.][]{Reynoso2008MNRAS,2008A&A...485..623R}. Only the impact of the jets on the environment would produce thermal X-rays and extended radio emission, but in large stellar associations there are many extended sources of uncertain origin. The protons would diffuse until they reach a suitable target. Then a gamma-ray source would appear with a significantly higher emissivity than the average gamma-ray production rate in the Galactic disk \citep{1996A&A...309..917A}. However, the origin of these PeV protons could be significantly displaced from the gamma-ray source itself, making identification difficult \citep[see][for a discussion of such a scenario]{1996A&A...309..917A}. A similar scenario was explored by \cite{2005A&A...432..609B}, where the authors investigated hadronic microquasars located in dense regions of the interstellar medium.

The ultra-relativistic protons ($E_p>1\,$PeV) accelerated by the reverse shock in the lobes of S26 are injected into the ISM with a total power of $\sim 10^{36}\,{\rm erg\,s^{-1}}$. As we just mentioned, they then diffuse and can reach regions of higher density, such as nearby molecular clouds. The energy distribution of protons at a distance $D$ and time $t$ can be determined by solving the diffusion equation \citep{1964ocr..book.....G}. In the spherically symmetric case, it reads
\begin{equation}
    \frac{\partial f}{\partial t}=\frac{D}{R^2}\frac{\partial}{\partial R}R^2 \frac{\partial f}{\partial R}+\frac{\partial}{\partial E}(Pf)+Q,
\end{equation}
where $f\equiv f(E,R,t)$ is the distribution function of particles, $P=-({\rm d}E/{\rm d}t)$ the continuous energy loss rate, $Q\equiv Q(E,R,t)$ the source function, and $D\equiv D(E)=D_0 E^{\delta}$ is the diffusion coefficient. \cite{Atoyan_etal_1995} found a solution to this equation, as a function of the diffusion radius, i.e., the radius of the sphere up to which the particles of energy $E_{\rm p}$ effectively propagate during the time $t$ after they left the acceleration site \citep{2005A&A...432..609B}. 

The fraction of cosmic rays that penetrate a cloud depends on a number of factors, including the magnetic field, the diffusion coefficient, advection, and energy losses due to ionization of neutral hydrogen atoms \citep{Cesarsky1978A&A,Morlino2015MNRAS,Dogiel2018ApJ,Ivlev2018ApJ}. Approximately $\sim 0.1$ of the proton population can produce gamma-ray emission through neutral pion decay if they interact with a nearby target. This, in turn, results in the conversion of up to 10\% of the injected power into gamma rays if the cloud is dense enough. This means an emission of $\sim 10^{34}\,{\rm erg\,s^{-1}}$ of very energetic gamma rays. If one of these sources is located in our Galaxy, approximately 10 kpc away, the flux on Earth would be about $\sim10^{-12}\,{\rm erg\,cm^{-2}\, s^{-1}}$, which is easily detectable by LHAASO. Thus, it is possible that undetected super-Eddington MQs could be PeVatrons behind some of the unidentified UHE sources. The problem of gamma rays from the interaction of CRs with molecular clouds was addressed by \cite{Aharonian_1991,Combi&Romero1995,Aharonian2001,Gabici_etal_2007}, among other authors. We plan to investigate this issue in detail elsewhere, following the approach of these papers, adapted to the case of super-Eddington MQs.

Other tracers that could help characterize the source could be neutrinos. The dominant scenario in the production of these particles will be \textit{pp} interactions. Although the total power in neutrinos will be greater than in the case of SS433 \citep[see e.g.,][]{Reynoso2008MNRAS}, S26 is thousands of times farther away, so the number of neutrinos reaching Earth will be lower. Furthermore, the only neutrino observatory currently operating at these energies is IceCube, which is located at the South Pole but detects neutrinos coming from the North, and S26 is only observable from the Southern Hemisphere.

As for the total power in CRs injected by a microquasar similar to S26 in the host galaxy, although it is large, of the order of $5\times 10^{38}\,{\rm erg\,s^{-1}}$ for $E_p>1\,$GeV (i.e, $\sim 5\%$ of the CR power in our Galaxy), the lifetime of the super-Eddington phase ($\sim10^5\,$yr) is not long enough to make a strong contribution to the average cosmic ray energy density. The contribution of a single source would be diluted by diffusion and cooling in the CR sea of the galaxy.

We plan to continue the detailed study of cosmic ray production in super-Eddington MQs and their propagation and interaction in the future.

\section{Conclusions} \label{sect: conclusions}

We have developed a model to explain the puzzling features of the extragalactic microquasar S26, which has the most powerful jets yet observed in accreting binaries. We proposed that the four-order-of-magnitude difference between the jet and apparent disk luminosities in S26 is caused by the complete absorption of the disk radiation by the wind ejected from the super-Eddington disk. We suggest that some of the energy driving the jet must be extracted from the ergosphere of a high-spin black hole. 

We have also reduced and analyzed data from multi-epoch X-ray observations of the X-ray \textit{XMM-Newton} space telescope that contain S26 in the field of view. The information obtained complements that already available from \textit{Chandra} observations at higher resolution. We found that the \textit{XMM-Newton} can be well fitted by a thermal blackbody spectrum with a peak temperature of $\sim 0.1$ keV and a hard power-law tail of index $\Gamma\sim 1.5$, with large uncertainties. We cannot rule out variability in this component.  In our model, we propose that the power-law component is the result of particle acceleration by internal shocks formed near the base of the jet, just above the wind photosphere, in the region where the outflow begins to be matter-dominated. The X-rays are produced by the electron synchrotron mechanism. The blackbody component, on the other hand, is produced in the terminal region of the jets, as a thermal contribution from the forward shock formed at the hotspots. The radio emission observed with ATCA can be explained as electron-synchrotron radiation from electrons accelerated by the reverse adiabatic shock in the lobes. All this is in agreement with the available X-ray \textit{Chandra} images and the ATCA radio maps obtained by \cite{2010Natur.466..209P} and \cite{soria2010}.

We calculate that the protons accelerated by the shocks at both acceleration sites reach energies greater than $\sim 1\,$PeV. Therefore, we propose that S26 can be considered a PeVatron microquasar with a super-Eddington accreting black hole. A similar but weaker microquasar in our Galaxy could also produce very high energy particles. If it is hidden within a stellar association, it could inject significant power into CRs that could ``illuminate'' molecular clouds and other matter overdensities, giving rise to UHE gamma-ray sources such as those observed by LHAASO. The scenario is similar to that proposed long ago by \cite{1996A&A...309..917A}.

\begin{acknowledgements}
The authors thank the Editor Steven Shore and the anonymous referee for a careful and constructive review, and for his/her comments, which have improved this paper. 
This work was supported by grant PIP 0554 (CONICET). LA acknowledges the Universidad Nacional de La Plata for the education received, Pablo Sotomayor for useful comments, and the 8th HEPRO organizers. GER was funded by  PID2022-136828NB-C41/AEI/10.13039/501100011033/ and through the ``Unit of Excellence María de Maeztu 2020-2023'' award to the Institute of Cosmos Sciences (CEX2019-000918-M).
\end{acknowledgements}

\bibliographystyle{aa} 
\bibliography{main} 

\end{document}